\documentclass[fleqn,usenatbib]{mnras}

\usepackage{newtxtext,newtxmath}
\usepackage[T1]{fontenc}

\DeclareRobustCommand{\VAN}[3]{#2}
\let\VANthebibliography\thebibliography
\def\thebibliography{\DeclareRobustCommand{\VAN}[3]{##3}\VANthebibliography}

\usepackage{graphicx}	% Including figure files
\usepackage{amsmath}	% Advanced maths commands
\usepackage{chemformula}

\defcitealias{Heger2010}{HW10}  %\citetalias{Heger2010} or \citepalias{Heger2010}
\defcitealias{Limongi2018}{LC18}
\defcitealias{Woosley1995}{WW95}
\defcitealias{Hartwig2018}{Har18}
\usepackage[normalem]{ulem}
\usepackage{xcolor}
\usepackage{comment}

\newcommand{\Zsun}{Z_{\odot}}

\newcommand{\reply}{}

\title[The descendants of Pop III stars]{Characterising the true descendants of the first stars}

\author[I. Vanni et al.]{
Irene Vanni,$^{1, 2}$\thanks{E-mail: irene.vanni@unifi.it}
Stefania Salvadori,$^{1, 2}$
Ása Skúladóttir,$^{1, 2}$
Martina Rossi,$^{1, 2}$
and Ioanna Koutsouridou $^{1, 2}$
\\
$^{1}$Dipartimento di Fisica e Astrofisica, Università degli Studi di Firenze, Via G. Sansone 1, I-50019, Sesto Fiorentino, Italy\\
$^{2}$INAF/Osservatorio Astrofisico di Arcetri, Largo E.Fermi 5, I-50125, Firenze, Italy
}

\date{Accepted 2023 September 20. Received 2023 August 10; in original form 2023 June 14}

\pubyear{2023}

\begin{document}
\label{firstpage}
\pagerange{\pageref{firstpage}--\pageref{lastpage}}
\maketitle

\begin{abstract}
The metal-poor stars in the Galactic halo are thought to show the imprints of the first (Pop~III) stars, and thus provide a glance at the first episodes of star formation. In this work, we aim at understanding whether all very metal-poor stars formed in environments polluted by Pop~III supernovae (SNe) and at what level. With a general parametric model for early metal enrichment, we study the chemical abundances (from C to Zn) of an environment imprinted by a single Pop~III SN. We investigate how these abundances depend on the initial mass and internal mixing of Pop~III stars, as well as on their SN explosion energy. We then study how subsequent generations of normal (Pop~II) SNe affect the Pop~III chemical signatures.
By comparing the observed chemical abundances with our model predictions, we show that stars with [C/Fe]~$>+2.5$ form in environments polluted purely by low-energy Pop~III SNe ($E_{\rm SN}<2\times 10^{51}$~erg). At lower [C/Fe], stars can be imprinted either by Pop~III only, or also by normal Pop~II SNe. The probability of being enriched by Pop~II SNe increases as [C/Fe] decreases. When Pop~II stars contribute more to the pollution, they wash out the diverse chemical peculiarities left by the different Pop~III SNe, and the chemical dispersion between their descendants decreases. We conclude that C-normal stars ($\rm [C/Fe] \leq +0.7$) have likely been enriched by Pop~II SNe at a $\geq 50\%$ level and we identify in the abundance scatter a key diagnostic to pinpoint the signature of Pop~III SNe.

\end{abstract}


%%%%%%%%%%%%%%%%%%%%%%%%%%%%%%%%%%%%%%%%%%%%%%%%%%

%%%%%%%%%%%%%%%%% BODY OF PAPER %%%%%%%%%%%%%%%%%%

\section{Introduction}

\label{sec:intro}

The first (Pop~III) stars are believed to form about 200-400 Myrs after the Big Bang (at redshift z $ \sim 30-20$) in primordial composition gas clouds dwelling in low-mass ($M \sim 10^6 \mathrm M_{\odot}$) dark matter structures called mini-halos \citep[e.g.][]{Tegmark1997,Abel2002,Yoshida2003,Bromm2013,Glover2013,Greif2015}. Given the lack of heavy elements (i.e. metals) and the low virial temperature of the mini-halos ($T_{\rm vir} \le 10^4$ K), the only available channel to cool down the gas is the molecular hydrogen (H$_2$), which is not an effective coolant and refrigerates the collapsing gas clouds to higher central temperatures \citep[$T_{\rm c} \sim 200$ K, see][]{Bromm2011} than what is seen in present-day star-forming metal-rich clouds ($T_{\rm c} \sim 10$ K). The higher primordial gas temperature leads to both more massive proto-stellar gas clouds \citep[$M \sim 1000 \mathrm M_{\odot}$, e.g. see][]{Abel2002,Yoshida2003} and higher gas accretion rates onto the proto-stars \citep[e.g.][]{OmukaiPalla2003,Hosokawa2009a,Ohkubo2009}. These results, which only depend on the lack of heavy elements, are very robust and suggest that stars formed in primordial composition environments (i.e. Pop III stars) are more massive than present-day normal (Pop~II/I) stars \citep[e.g.][]{Hosokawa2011,Susa2014,Hirano2014}. 

Hydrodynamic cosmological simulations \citep[see e.g.][]{Hosokawa2011,Hirano2015,Greif2015,Skinner2020} and indirect studies \citep[e.g.][]{Tumlinson2006,Ballero2006,Salvadori2007,Hartwig2015,Salvadori2015,Ma2017,Sarmento2017,Salvadori2019,Rossi2023} have provided some hints on the still debated mass range and distribution of Pop~III stars. 
Many factors play a role in the fragmentation and star-formation processes, like turbulence \citep[e.g.][]{Greif2011} or radiative feedback from the protostar \citep[e.g.][]{Hosokawa2011}, and, depending on their assumptions, cosmological simulations often find different results. In conclusion, Pop~III stars can form either isolated (if massive) or in very small groups (if sub-solar) and their predicted initial masses range between $0.1$ and $1000\, \mathrm M_{\odot}$, with a characteristic mass probably $\gtrsim 10 \,\mathrm M_{\odot}$ \citep[e.g.][]{Hirano2014,Susa2014,Hirano2017,DeBennassuti2017,Sarmento2019,Rossi2021}. Therefore, the majority of Pop~III stars are short-lived, with lifetimes of a few Myrs.

Pop~III stars with masses in the range [10;~100]\,$\mathrm M_{\odot}$ end their lives as supernovae (SNe), exploding with a variety of energies \citep[e.g.][]{Kobayashi2006,Heger2010}; and stars in the range [140; 260]\,$\mathrm M_{\odot}$ explode as Pair Instability Supernovae \citep[PISN, see][]{Heger2002a}, both enriching the interstellar medium (ISM) with newly produced metals. The abundances of the different chemical elements depend on the properties of the primordial star, i.e.~the initial mass and the SN explosion energy. When the total metallicity of the ISM overcomes the critical metallicity ($\rm Z_{cr} \sim 10^{-5\pm 1} \mathrm{Z_{\odot}}$; e.g., \citealp{Omukai2000,Bromm2001,Maio2010,Schneider2010}) the gas efficiently cools down and normal Pop~II stars form, with lower masses in the range [0.1; 100]\,$\mathrm M_{\odot}$. Since stars with masses $< 0.8\, \mathrm M_{\odot}$ have lifetimes longer than the age of the Universe, \emph{second generation} (2G) stars might have survived and be observable today as pure descendants of Pop~III stars. In principle, their photospheres should represent the chemical composition of the material ejected from the first SNe. The study of these old fossils in order to retrieve information about the properties of the first stars is called \emph{stellar archaeology}. 

The Milky Way stellar halo is its most metal-poor component and it has been intensively studied during the last 20 years \citep[e.g.][]{Beers1992,Cayrel2004a,Yong2013a,Roederer2014,Bonifacio2021}: here the most metal-poor \citep[][]{Caffau2011a}, iron-poor \citep[][]{Keller} and old ($\tau \sim 13 \times 10^{9}$ yrs, see \citealp{Cowan2002}) stars have been observed. The stars in the Galactic halo have iron abundances\footnote{[A/B]=$\log \left( \dfrac{N_{\rm A}}{N_{\rm B}} \right) - \log \left( \dfrac{N_{\rm A}}{N_{\rm B}} \right)_{\odot} $}, [Fe/H], that span more than 6 orders of magnitude, from less than -7 to almost 0 dex. Stars with [Fe/H] $\le -1$ are called metal-poor (MP) stars. They can be divided into two categories depending on their carbon-to-iron elemental abundances. 

The C-normal stars, [C/Fe]$ \le 0.7$, have relatively homogeneous chemical abundance patterns \citep[e.g.][]{Cayrel2004a,Yong2013a}. 
The MP stars with [C/Fe]$ > +0.7$ are called Carbon-Enhanced Metal-Poor (CEMP) stars and are again sub-categorized depending on the abundances of the elements produced in slow (s-), or rapid (r-) neutron capture processes \citep[see][]{Beers2005}. The stars with overabundance of these elements, like barium, are called CEMP-s(/r) ([Ba/Fe] $> 1$), while the other are called CEMP-no stars ([Ba/Fe] $< 0$). The presence of s-process elements is consistent, and confirmed by observations, with the scenario where the star is, or has been, in a binary system with an Asymptotic Giant Branch star, which transferred mass \citep[see][]{Abate2015} and altered the chemical composition of the photosphere. Therefore, the C-enhancement measured in the photospheres of CEMP-s(/r) stars has been acquired during their lifetimes.
On the other hand, the observed CEMP-no stars are not primarily in binary systems \citep[see][]{Starkenburg2014,Arentsen2019} and their high values of $\mathrm {^{12}C/^{13}C}$ show that their surface composition has not been altered by mass transfer \citep[see][]{Aguado2022}. Hence, evidence suggests that CEMP-no stars most likely inherited their C-enhancement from their birth clouds.

Both the fraction of CEMP-no stars and their [C/Fe] ratio increase towards low [Fe/H], denoting a likely connection to the descendants of the first stars. Indeed, the chemical abundances of the most iron-poor CEMP-no halo stars are consistent with being enriched by a single intermediate-mass ($M \gtrsim 20\, \mathrm M_{\odot}$) Pop~III SN exploding with low-energy, the so-called faint SNe ($E_{\rm SN} < 10^{51}$ erg), experiencing mixing and fallback during the explosion \citep[e.g.][]{Umeda2003,Iwamoto2005a,Marassi2014a}. The CEMP-no halo stars at higher [Fe/H], instead, appear to be consistent with being polluted either by only Pop~III stars exploding with different energies (see \citealp{Welsh2020}) or by a combination of Pop~III SNe and normal Pop~II stars \citep[see][]{DeBennassuti2017,Koutsouridou2023}. In conclusion, many theoretical works find that CEMP-no halo stars are linked to Pop~III stars. However, it is still unclear whether they are all pure descendants of the first stars or if they are also contaminated by Pop~II stars. Eventually, what fraction of the metals in the birth environment of CEMP-no stars has been contributed by Pop~II stars?

Only recently the imprints of a single very energetic Pop~III SN, the so-called hypernovae ($E_{\rm SN} > 5 \cdot 10^{51}$ erg), have been detected in C-normal metal-poor stars residing in the Galactic halo \citep[][]{Placco2021} and in the dwarf galaxy Sculptor \citep[][]{Skuladottir2021,Skuladottir2023}. In addition, stars with signs of an enrichment by an extremely energetic Pair Instability Supernova (PISN, $E_{\rm SN} > 10^{52}$ erg) combined with normal Pop~II SNe have been found in the Galactic halo \citep[][]{Aoki2014,Salvadori2019}. Contrary to faint SNe, the true descendants of very energetic Pop~III SNe seem extremely rare. Can they only be found at low [C/Fe]? Finally, are all metal-poor halo stars predominantly polluted by Pop~III stars? 

The aim of this paper is to answer these questions and chemically characterize the descendants of the first stars. In order to achieve these aims we exploit the chemical abundances of {\it all} the available elements for CEMP-no and C-normal metal-poor halo stars and interpret their different abundance patterns with new theoretical models. This has never been explored before since previous theoretical studies only focused on specific chemical abundances.

\section{Very metal-poor stars in the Galactic halo}

\label{data}

In this Section we analyse the chemical composition of very metal-poor stars (VMP; [Fe/H] $ < -2$) observed in the Galactic halo, for which high-resolution observations ($R=\dfrac{\Delta\lambda}{\lambda} > 30\,000$) are available. This sample of more than 200 stars includes the 190 and 35 metal-poor stars from \citet{Yong2013a} and \citet{Cayrel2004a}, respectively, and additional 25 stars with [Fe/H] $\le -3$, among which 21 have [Fe/H] $\le -4$, from: \citet{Christlieb2004}; \citet{Norris2007}; \citet{Caffau2011a}; \citet{Hansen2014}; \citet{Keller}; \citet{Frebel2015a}; \citet{Bonifacio2015a}; \citet{Li2015}; \citet{Bonifacio2018}; \citet{Francois2018}; \citet{Starkenburg2018}; \citet{Aguado2019}; \citet{Ezzeddine2019}. The stellar abundances for the majority of these stars are not corrected for 3D and non-LTE effects, and this will be discussed in Sect.~\ref{discussion}. We exclude CEMP-s(/r) stars, i.e. with [C/Fe] $ > +0.7$ and [Ba/Fe] $ > 0$, whose abundances are not representative of their birth environment (see Sect.~\ref{sec:intro}). We end up with the chemical abundances, of 16 elements in total (including the iron abundance), of 132 stars: for all of them we have the iron and carbon abundances, but measurements for other elemental abundances are available only for some of them (\reply{see Figs. \ref{fig:histo_colors} and \ref{fig:VMP_mean}}). 
We have corrected the carbon abundances of all the stars in this sample to account for the effect of evolutionary status by exploiting the online tool\footnote{http://vplacco.pythonanywhere.com/} presented in \citealt{Placco2014c}.

\subsection{The iron and carbon abundances}

\begin{figure}
    \centering
    \includegraphics[width=\linewidth]{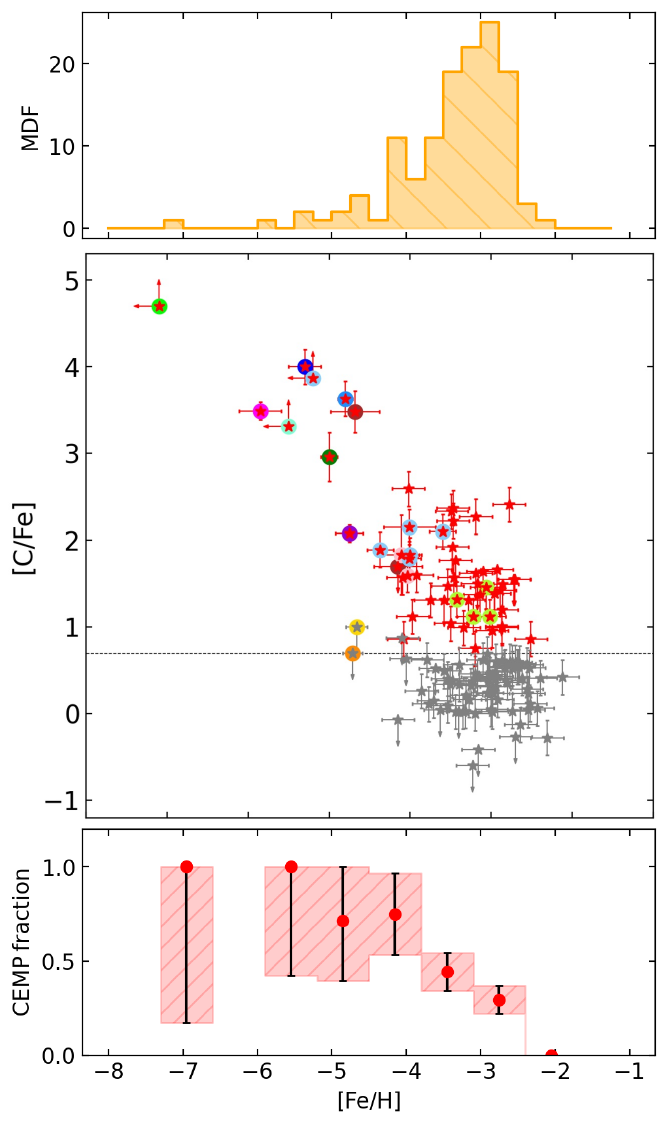}
    \caption{Properties of our literature sample (Sect.~\ref{data}). \emph{Top panel}: Metallicity Distribution Function (MDF). \emph{Middle panel}: [C/Fe] vs [Fe/H] for CEMP-no (red) and C-normal (grey) stars, corrected for the evolutionary status \citep[][]{Placco2014c}. \reply{The stars have circles color-coded with their literature sources (if they don't have circles, they pertain to \citealp{Yong2013a}'s sample): \citet[][blue]{Christlieb2004}; \citet[][purple]{Norris2007}; \citet[][orange]{Caffau2011a}; \citet[][brown]{Hansen2014}; \citet[][medium green]{Keller}; \citet[][dark green]{Frebel2015a}; \citet[][darker cyan]{Bonifacio2015a}; \citet[][pink]{Li2015}; \citet[][medium cyan]{Bonifacio2018}; \citet[][light green]{Francois2018}; \citet[][yellow]{Starkenburg2018}; \citet[][light cyan]{Aguado2019}; \citet[][fuchsia]{Ezzeddine2019}}. The observational errors are shown, and the upper/lower limits are presented with arrows.
    \emph{Bottom panel}: CEMP-no fraction, with the poissonian errors. 
    }
    \label{fig:CEMPeCno}
\end{figure}

The carbon abundance is one of the most important and most extensively exploited diagnostics in stellar archaeology \citep[e.g.][]{Cooke2014,Salvadori2015,DeBennassuti2017,Hartwig2018,Chiaki2020,Liu2021,Rossi2023}. In Fig.~\ref{fig:CEMPeCno}, we present the main properties of our literature sample. The metallicity distribution function (MDF; top panel) peaks at [Fe/H] $\sim -3$, but extends down to [Fe/H] $< -7$ (\citealt{Keller}: [Fe/H] $< -7.1$ with 3D corrections). In the middle panel, we show [C/Fe] with respect to [Fe/H], with observational errors or upper/lower limits, distinguishing between C-normal ([C/Fe] $ \le +0.7$) and CEMP-no stars ([C/Fe] $> +0.7$). \reply{The stars that come from literature sources different from \citealt{Yong2013a} are color-coded with their sources}. Two ultra metal-poor stars ($\rm[Fe/H]<-4$) only have upper limits of [C/Fe] $ < +0.7$ and $< +1.0$ with 3D corrections \citep{Caffau2011a,Starkenburg2018}, we thus consider them as C-normal stars. The [C/Fe] abundance ratios increase as [Fe/H] decreases, and all stars at [Fe/H] $\lesssim -5$ are C-enhanced. In the bottom panel of Fig.~\ref{fig:CEMPeCno} we present the CEMP fraction, with the poissonian error computed as $\sqrt{N_{\rm CEMP}}/N$, where N is the total number of stars and $N_{\rm CEMP}$ the number of CEMP-no stars in each [Fe/H] bin. 
In the cases where we have less than five stars in the bin (i.e. at the lowest [Fe/H]) we use the values reported in \citet{Gehrels1986}, and if the fraction is 0 we don't show the error bars. We see in Fig.~\ref{fig:CEMPeCno} that the CEMP fraction increases strongly with decreasing [Fe/H]: it is around 1 for [Fe/H] $\lesssim -5$, 0.7 for [Fe/H] $\sim -4$, and 0.3 for [Fe/H] $ \sim -3$, which is consistent with other studies \citep[e.g.][]{Placco2014c,Arentsen2022}. 

\subsection{The complete abundance pattern}

\begin{figure}
    \centering
    \includegraphics[width=\linewidth]{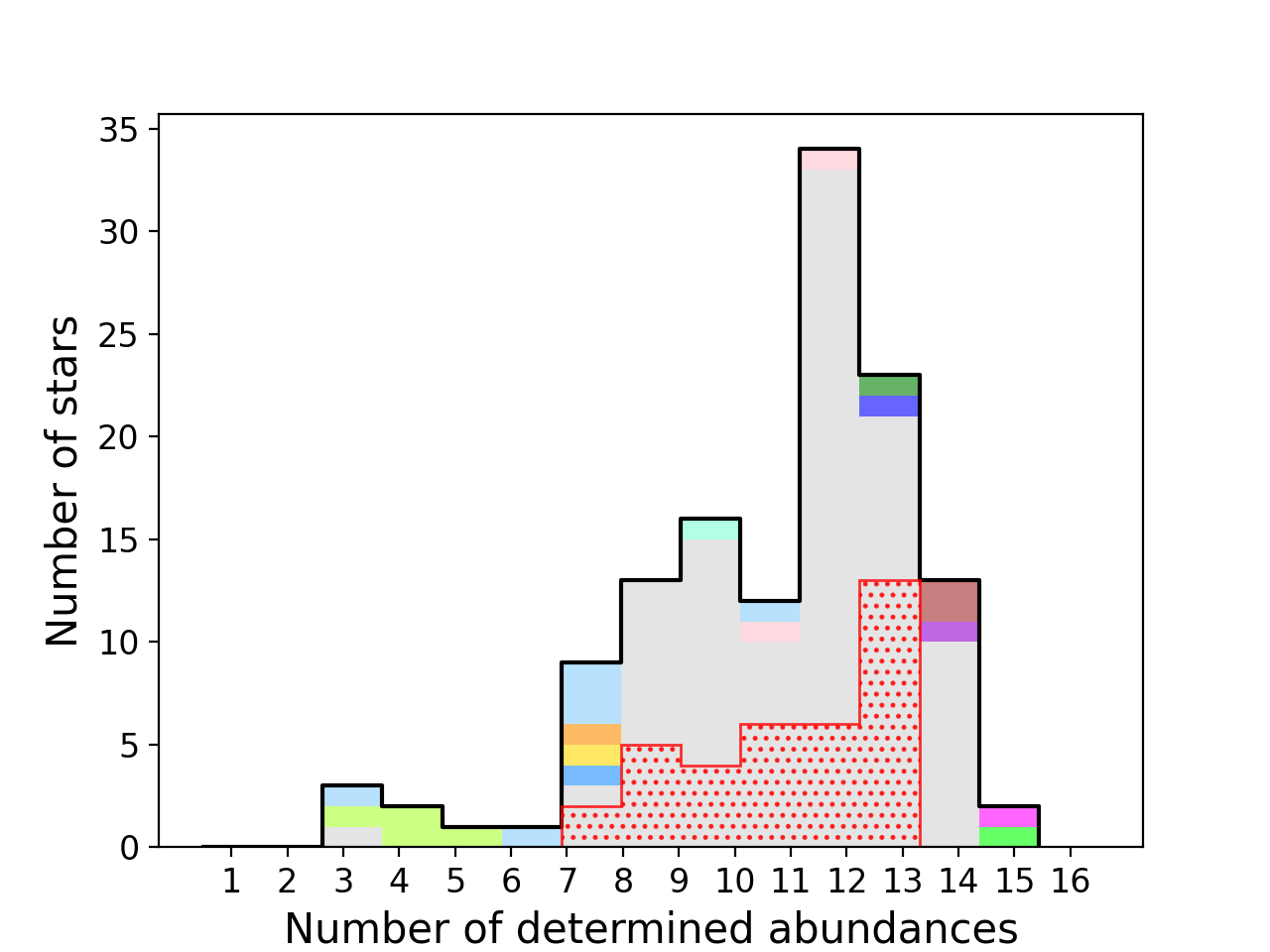}
    \caption{\reply{Number of available chemical abundances for the stars of our literature sample. The histogram is color-coded with the literature source of each star: grey for all the stars from \citet{Yong2013a}, whose CEMP-no stars are highlighted with the red area, and the other colors are the 21 additional sources as in Fig.~\ref{fig:CEMPeCno}.}}
    \label{fig:histo_colors}
\end{figure}

\begin{figure*}
    \centering
    \includegraphics[width=\linewidth]{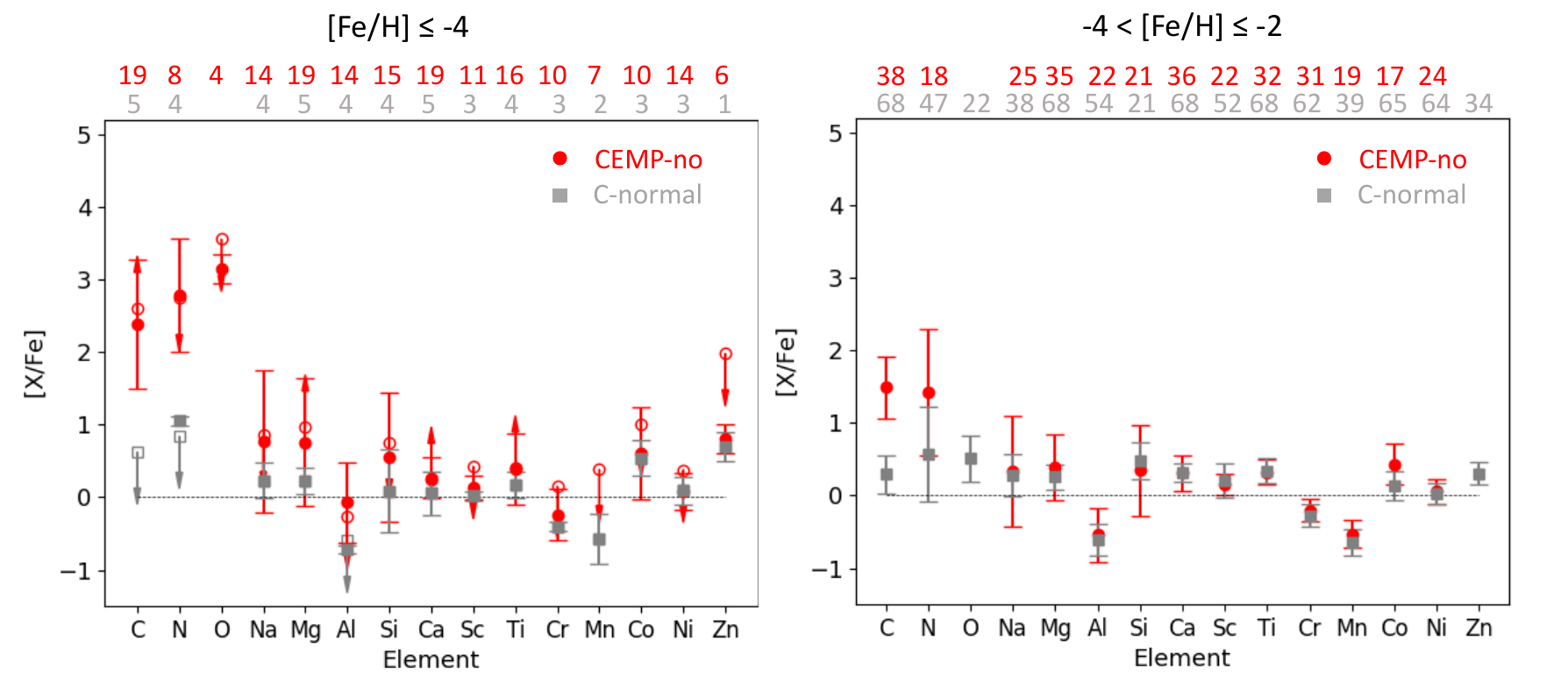}
    \caption{Mean abundance patterns for the stars in our literature literature sample (see text for the references) for CEMP-no (red points) and C-normal (grey points) stars, for [Fe/H] $\le -4$ (left), and $\in (-4; -2]$ (right). On top are written the number of CEMP-no (red) and C-normal (grey) stars included for each chemical element. Filled points exclude limits, while open points include upper/lower limits. The error bar is the standard deviation of measurements, when there is only one observation we use the observational error. Measurements that present upper limits in both [Fe/H] and [X/H] are not included. The arrows point down if [X/Fe] has more upper than lower limits, and vice versa.  
    }
    \label{fig:VMP_mean}
\end{figure*}

In addition to the carbon abundances, we also compared the measured abundances of other chemical elements to those predicted by our model. \reply{In Fig.~\ref{fig:histo_colors} we show how many abundances are reported for each star of our literature sample, color-coded with the source. The average number of chemical abundances provided per star is 11, but 12 abundances is the most numerous. None of the stars in this literature sample have all the 15 chemical abundances, but for all of them we have at least 3: iron, carbon and another element. The most reported chemical abundances, apart from carbon and iron, are magnesium, calcium and titanium. In addition to the abundances shown in Fig~\ref{fig:histo_colors}, we take O and Zn separately from the stellar sample of \citet{Cayrel2004a}, but these are not counted in the histogram.} 

In Fig.~\ref{fig:VMP_mean}, we show the mean chemical abundance ratios with respect to iron, [X/Fe], for the CEMP-no (red) and C-normal (grey) stars of our literature sample, distinguishing between [Fe/H] $\le -4$ (left) and [Fe/H]$\in (-4; -2]$ (right). In the case of [Fe/H] $\le -4$, we include both abundance averages excluding (filled points, error bars) and including (open symbols, arrows) upper/lower limits on [X/Fe]. 
When [X/Fe] has more upper than lower limits we put an arrow pointing down (or vice versa). We never include measurements that only have upper limits in both [Fe/H] and [X/H]. For [Fe/H] $\in (-4; -2]$ we have at least 18 measured abundances for both C-normal and CEMP-no stars for all the chemical elements, therefore no upper/lower limits are included. For the \citet{Cayrel2004a} sample, only stars are included that have not undergone internal mixing of C, according to \citet{Spite2005}. \citet{Yong2013a} do not provide the abundances [O/Fe] and [Zn/Fe], thus they are missing for the CEMP-no stars in Fig.~\ref{fig:VMP_mean}.

We first notice from Fig.~\ref{fig:VMP_mean} that CEMP-no stars have mean C, N and O abundance ratios that are typically much higher than those of C-normal stars. However, we should point out that more than 60\% of the measured [N/Fe] and [O/Fe] for the CEMP-no stars are upper limits, hence their real values can be considerably lower than the ones represented here. Furthermore, at [Fe/H]$<-4$, we see that also the mean abundance ratios of many alpha-elements (e.g. Mg, Al, Si) over iron is higher in CEMP-no stars than in C-normal stars, while heavier elements (Co, Ni, Zn) are comparable between CEMP-no and C-normal stars. Finally, at $-4<$[Fe/H]$<-2$,  we see that, with the only exception of C and N, the mean abundance ratios of CEMP-no and C-normal stars are consistent among each other. Still, we notice that CEMP-no stars typically have larger star-to-star scatter\footnote{Here the star-to-star scatter is quantified with the standard deviation among different measurements} than C-normal stars, even when the number of observed stars is similar or of the same order of magnitude (e.g. Na, Mg, Si, Al). 

\reply{We made several checks to ensure that the scatter is not driven by systematic errors of the different literature sources. For [Fe/H] $\in (-4; -2]$, the stars from \citet{Yong2013a} are more numerous than stars from other literature sources (eight). Thus the average abundance ratios and the standard deviations in this case are driven by the stars of \citet{Yong2013a}'s stellar sample. This assures that the scatter between the abundances of different stars is not because of the different literature sources and analysis methods, it is instead an intrinsic property of our stellar sample. We also emphasize that the scatter in abundance ratios at the lowest [Fe/H]$<-4$ is very large, often exceeding 1~dex, which is more than what is expected from different analysis methods. This is also confirmed by the uniform carbon-analysis of \citet{Norris2019} of the most metal-poor stars known.} 

In conclusion, the star-to-star scatter in the chemical abundance ratios of CEMP-no stars is large at $-4<$[Fe/H]$<-2$ and huge for [Fe/H]$<-4$ (see also Fig.~1 in \citealt{Vanni2023}). Conversely, C-normal stars show a small dispersion in their abundance ratios as already pointed out by \citet{Cayrel2004a}. For this reason, the star-to-star scatter can be used as a new path to understand which stars polluted the birth environment of metal-poor halo stars. Indeed, a group of stars which show a small scatter in the abundance ratios, same as the C-normal stars, might have formed in an environment chemically enriched by multiple stellar populations. On the contrary, the stars that exhibit, one from each other, different chemical abundances might have been enriched by one or few SNe.

\section{The semi-analytical model}

\label{sec:model}
In order to characterise the descendants of Pop~III stars, we adopt the parametric model described and developed in \citet{Salvadori2019}. The model follows the chemical properties of primordial galaxies in which a Pop~III Pair Instability Supernova (PISN) exploded. Since no star polluted only by PISNe has been discovered to date, the authors aim at finding which of these chemical features are maintained after also Pop~II stars have contributed up to 50\% of the total amount of metals in the ISM.
Here we expand this model to study the chemical imprints of Pop~III SNe covering a larger range of masses and explosion energies.

\subsection{Model recap}
\label{sec:original}

We summarize here the principal features of the parametric study presented in \citet{Salvadori2019}.

\begin{enumerate}
    
\item {\it The first stellar generation.} The original model assumes that a single\footnote{Or more than one but all with the same mass and explosion energy.} Pop~III star with initial mass in the range M $\in[150; 260]\,\mathrm M_{\odot}$ forms in a primordial proto-galaxy. In this mass range the stars end their lives as energetic Pair Instability SN (PISN), $\mathrm E_{\rm SN} \in [10^{52}; 10^{53}]$ erg, which completely destroys the star. The yields adopted for these stars are the ones from \citet{Heger2002a}. The uncertainties linked to the first episodes of star formation are enclosed in two parameters: the star-formation efficiency, $f_*$, and the dilution factor, $f_{\rm dil}$. The star-formation efficiency is the fraction of the gas in the primordial galaxy which is converted in stars. The dilution factor is the ratio between the fraction of the initial gas mass in which the metals are diluted (which can also be $> 1$ if, e.g., primordial gas is infalling in the galaxy) and the fraction of the newly produced metals that are effectively retained by the galaxy. In simple terms, the dilution factor quantifies how much the metals are diluted in the hydrogen gas. 
\\
\item {\it The chemical abundances of the Pop~III-polluted gas.} With just one star exploding per primordial proto-galaxy, the yield of a certain element, $Y_{\rm X}^{\rm III}$, is simply the ratio between the mass of this element X ejected by the star and the initial mass of the star. The abundance of a chemical element X with respect to hydrogen in the ISM of the hosting primordial halo, after the first generation of stars polluted the gas, can be expressed as 
\begin{equation}
	\label{XsuHiii}
	\left[{\rm \dfrac{X}{H}}\right]_{\rm ISM}=\log\left[\dfrac{f_*}{f_{\rm dil}}\,Y_{\rm X}^{\rm III}\right]-\log\left[\dfrac{M_{\rm X, \odot}}{M_{\rm H, \odot}}\right],
\end{equation}

\noindent $M_{X, \odot}$ ($M_{H, \odot}$) is the mass of the element X (hydrogen) in the photosphere of the Sun. 
\\
\item {\it Normal Pop~II stars contribution.} After the pollution from a Pop~III star, the ISM metallicity usually overcomes the critical value, $Z_{\rm cr} \sim 10^{-5\pm 1} Z_{\odot}$, and normal Pop~II stars form with masses in the range $[0.1; 100]\,\mathrm M_{\odot}$. Here we use the critical metallicity $Z_{\rm cr}=10^{-4.5}\Zsun$ \citep[see][]{DeBennassuti2017}. After $\sim 3$ Myr from their birth, normal Pop~II core-collapse SNe start contaminating the ISM with newly synthesized metals, and in $\approx 30$~Myr all Pop~II SNe contribute to the ISM enrichment.
The total yields of Pop~II stars, $Y_{\rm X}^{\rm II}$, are thus computed by integrating over a Larson IMF\footnote{$\phi (m_\star) \propto m_\star^{-2.35} \times \exp{\left(-\dfrac{m_{\rm ch}}{m_\star}\right)}$} with $m_{ch}=0.35\,\mathrm M_{\odot}$ in the range $[0.1; 100]\,\mathrm M_{\odot}$ and adopting the yields from \citet{Woosley1995}. \reply{We consider only the stars that pollute the gas on short time scales ($\rm \lesssim 30 \: Myrs$), which is equivalent to assuming that the yields of Pop~II stars with $M_* \lesssim \rm 10 \:M_{\odot}$ are zero.}
By quantifying the amount of metals ejected by Pop~III stars with respect to the total metal's amount in the ISM through the free parameter ${f_{\rm PopIII}}$, 
we can compute the chemical abundances of the ISM after the contribution of both Pop~III and Pop~II SNe as:
\begin{equation}
	\label{XsuH_II}	{\rm[X/H]}_{\rm ISM}=\log\left[\frac{f_*}{f_{\rm dil}}\left(Y_{\rm X}^{\rm III}+\beta\dfrac{Y_{\rm X}^{\rm II}Y_Z^{\rm III}}{Y_Z^{\rm II}}\right) \right]-\log\left[\frac{M_{\rm X, \odot}}{M_{\rm H, \odot}}\right],
\end{equation}
where ${\rm \beta=(1-f_{\rm PopIII})/f_{\rm PopIII}}$. Note that [X/H] is affected by $f_*/f_{\rm dil}$, $f_{\rm PopIII}$, and the yields of Pop~III and Pop~II stars, while the abundances ratios, [X/Y], only depend on $f_{\rm PopIII}$ and the yields.
\\
\item {\it The parameter space.} The model has three free parameters: the star-formation efficiency, $f_*$, the dilution factor, $f_{\rm dil}$, and the fraction of metals contributed by Pop~III stars, $f_{\rm PopIII}$. From different physical arguments (see \citealp{Salvadori2019} for details) the range of $f_*$ can be assumed\footnote{Its minimum values are related to the minimum mass required to form stars in the coolest first star-forming mini-halos, its maximum values correspond to the more star-forming Ly$\alpha$-cooling halos.} to be $[10^{-4}; 10^{-1}]$, while $f_{\rm dil} \in [0.02; 10]$. The star-formation efficiency and the dilution factor are dependent on one another and their ratio, $f_*/f_{\rm dil}$, ranges in $[10^{-4}; 10^{-1}]$; we assume that all values within this range are equally probable.
To ensure a predominant contribution from Pop~III stars to the chemical enrichment the model explores the values ${\rm f_{\rm PopIII}=(0.5-1.0)}$.

\end{enumerate}

\subsection{Model implementation}

\subsubsection{Pop~III mass range and explosion energies}

\label{sec:PopIII_expl}

\begin{figure}
    \centering
    \includegraphics[width=\linewidth]{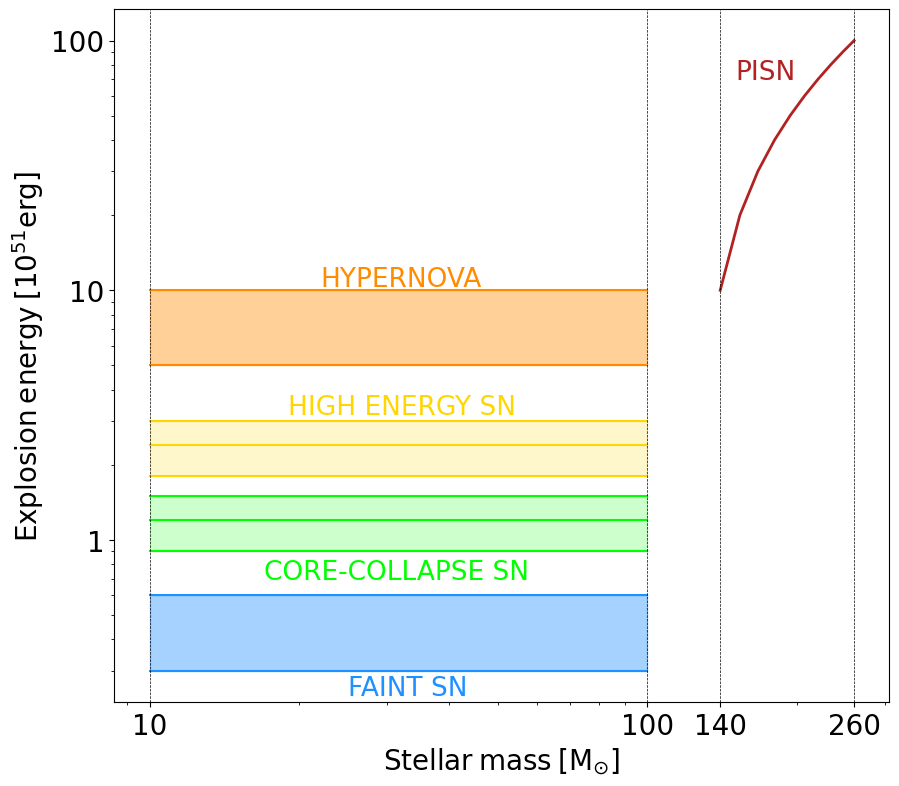}
    \caption{The extended mass range for Pop III stars and the corresponding energies as used in the implemented model.}
    \label{fig:energies}
\end{figure}

We extend the Pop~III mass range to [0.8;1000]$\,\mathrm M_{\odot}$ \citep[see][for the constraints on the minimum mass of Pop~III stars]{Rossi2021} in order to account for all the possible (initial) masses, and assume that Pop~III stars of different masses have equal probability to form. Since we are interested in understanding the chemical properties of the environment primarily polluted by Pop~III stars, here we only consider the contribution of stars with masses $M \ge 10\,\mathrm M_{\odot}$, which have lifetimes $\tau < 30$ Myr and quickly enrich the interstellar medium if exploding as SNe. \reply{Stars with masses up to $\rm \sim 8\,M_{\odot}$ (AGB stars) typically contribute to the chemical enrichment on timescales longer than hundreds of Myrs. Therefore their pollution is negligible during the first $\sim$30\,Myr of star formation.}

We use the yields of \citet{Heger2010}, which provide the metal yields for Pop~III stars with initial masses between $10$ and $100\,\mathrm M_{\odot}$, explosion energies, $E_{\rm SN}$, between $0.3 \times 10^{51}$ and $10 \times 10^{51}$\,erg, and different internal mixing efficiencies, $f_{\rm mix}$. In Fig.~\ref{fig:energies} we visualise the relationship between the explosion energy and the initial Pop~III stellar masses. In the range $10-100\,\mathrm M_{\odot}$, stars can be divided into four classes: \emph{faint SNe} (blue) with $E_{\rm SN}=0.3-0.6 \times 10^{51}$\,erg; \emph{core-collapse SNe} (CC SN, green) with $E_{\rm SN}=0.9-1.5 \times 10^{51}$\,erg; \emph{high-energy SNe} (HE SN, yellow) with $E_{\rm SN}=1.8-3.0 \times 10^{51}$\,erg; and \emph{hypernovae} (orange) with $E_{\rm SN}=5.0-10.0 \times 10^{51}$\,erg. \reply{Examples of Pop~III yields with different explosion energies are shown in Sect.~\ref{sec:comparison_III_II}.}

In our model we adopt four values for the mixing efficiency, $f_{\rm mix}=0.039, 0.0631, 0.100, 0.158$, and a single representative explosion energy for each class:
$0.6$ for faint SNe, $1.2$ for CC SNe, $3.0$ for HE SNe, and $10.0$ for hypernovae, in units of $10^{51}$\,erg. All the adopted mixing efficiencies and explosion energies are assumed to be equally probable. With these ${f_{\rm mix}}$, the predicted [C/Fe] for a second-generation star imprinted by a single Pop~III low-energy SN is consistent with what proposed by \citet{Iwamoto2005a} and \citet{Keller}. Notice that for these massive Pop~III stars the explosion energy does not depend on the mass of the star (Fig.~\ref{fig:energies}). The stars which explode with low energies (faint and CC SNe) only succeed in expelling the most external layers which are mainly composed by the lighter elements, C, O and Mg. However, these layers also carry heavier elements, like Fe, which are moved outwards by the mixing acting before the internal layers fall back onto the remnant. Therefore, the mixing is fundamental for low-energy SNe to expel also some amounts of heavy metals: if we suppose $f_{\rm mix}=0$, which is the case where the layers are not mixed at all, the mass of iron ejected by a 25\,$\mathrm M_{\odot}$ star exploding as faint SN would be $\ll 10^{-7}\,\mathrm M_{\odot}$. Conversely, when $f_{\rm mix}=0.1$ the iron yielded is $Y_{\rm Fe}\approx 10^{-3}\,\mathrm M_{\odot}$. On the other hand, the yields of more energetic SN explosions are much less dependent on the mixing efficiency. See Sect.~\ref{sec:appendix_2} in the Appendix for an in-depth study of the effects of the mixing efficiency on the [C/Fe] abundance ratio.

In Fig.~\ref{fig:energies} we also show the explosion energies of PISNe, i.e., stars with initial masses $[140-260]\,\mathrm M_{\odot}$: in this case the explosion energy is proportional to the initial mass of the star and ranges between $10^{52}$ and $10^{53}$ erg. PISNe with initial masses between $150$ and $260\,\mathrm M_{\odot}$ were already included in the model of \citet{Salvadori2019}, but here we also include masses between 140 and $150\,\mathrm M_{\odot}$, which have the lowest explosion energies and produce a peculiar abundance pattern, with [C/Fe] $> +2$, in the polluted ISM. 

\subsubsection{Pop~II stellar yields}

\label{PopII_yields}

The computed nucleosynthetic yields can change significantly depending on the stellar evolution model, as demonstrated in \cite{Romano2010}, thus affecting the results of the chemical evolution models that employ them and generating an intrinsic uncertainty. \reply{A direct comparison between the yields adopted in this work is shown in Sect.~\ref{sec:comparison_III_II}.}

The model developed in \citet{Salvadori2019} used the yields of \citet{Woosley1995} \citepalias[hereafter][]{Woosley1995} for Pop~II stars, which have initial metallicitities $Z = (0, 10^{-4}, 10^{-2}, 10^{-1}, 1) \mathrm{Z_{\odot}}$ , initial masses between $11$ and $40\,\mathrm M_{\odot}$ and explosion energy of $\sim 10^{51}$ erg. For Pop~II stars with $m_\star > 40 \mathrm{M_{\odot}}$ \citetalias{Woosley1995} did not compute the yields since those are supposed to directly collapse in a black hole. 

To relieve the uncertainty due to the choice of the model, we also use the recommended set of yields (set R), without rotation, from \citet{Limongi2018} \citepalias[hereafter][]{Limongi2018}. These yields are computed for Pop~II stars with initial metallicities $Z = (2.31 \times 10^{-3}, 10^{-2}, 10^{-1},1) {\rm Z_{\odot}}$ and initial masses between $13$ and $120\,\mathrm M_{\odot}$. In this case, the stars with initial masses higher than $25\,\mathrm M_{\odot}$ end their lives collapsing into black holes. These very massive stars therefore only contribute to the chemical enrichment through stellar winds. 

To be consistent while using the yields from the two models and avoid extrapolation we only consider the contribution to the chemical enrichment of Pop~II stars with masses $\geq 13\,\mathrm M_{\odot}$, which pollute the ISM with SN explosions or stellar winds. In other words, we compute the total yields of an element X provided by Pop~II stars, $Y_{\rm X}^{\rm II}$, by integrating over the Larson IMF (see Sect.~\ref{sec:original}) between $m_\star = [13,100] \mathrm{M_{\odot}}$ for both \citetalias{Woosley1995} and \citetalias{Limongi2018} data sets. Finally, since the minimum metallicity of the \citetalias{Limongi2018} yields, $Z_{\rm min}=2.31 \times 10^{-3} {\rm Z_{\odot}}$, is larger than the critical value we use these yields for all Pop~II stars with initial metallicity, $Z_{\rm cr}< Z_* <Z_{\rm min}$.

\reply{Even though the yields of Pop~III and Pop~II SNe are handled differently, i.e. Pop~II SNe have their yields integrated while Pop~III SNe are treated singularly, the difference in Pop~III and Pop~II SNe enriched environments is intrinsic in their yields, as shown in Sect.~\ref{sec:comparison_III_II}.}

\section{Results}

\label{results}
Our main aim is to study the chemical enrichment of an ISM imprinted by the first stars, i.e. the natal environment of the descendants of Pop~III stars. Hereafter, {\it Pop~III star descendants} will refer to the low-mass long-lived Pop~II stars which formed in an ISM where at least 50\% of the metals have been contributed by Pop~III stars. As we will show, a major Pop~II contribution to the metal supply ($>$50\%) would completely wash out the chemical imprint of Pop~III stars.

\subsection{Birth environments of second generation stars}

\label{sec:PopII}

\begin{figure*}
    \centering
    \includegraphics[width=0.90\textwidth]{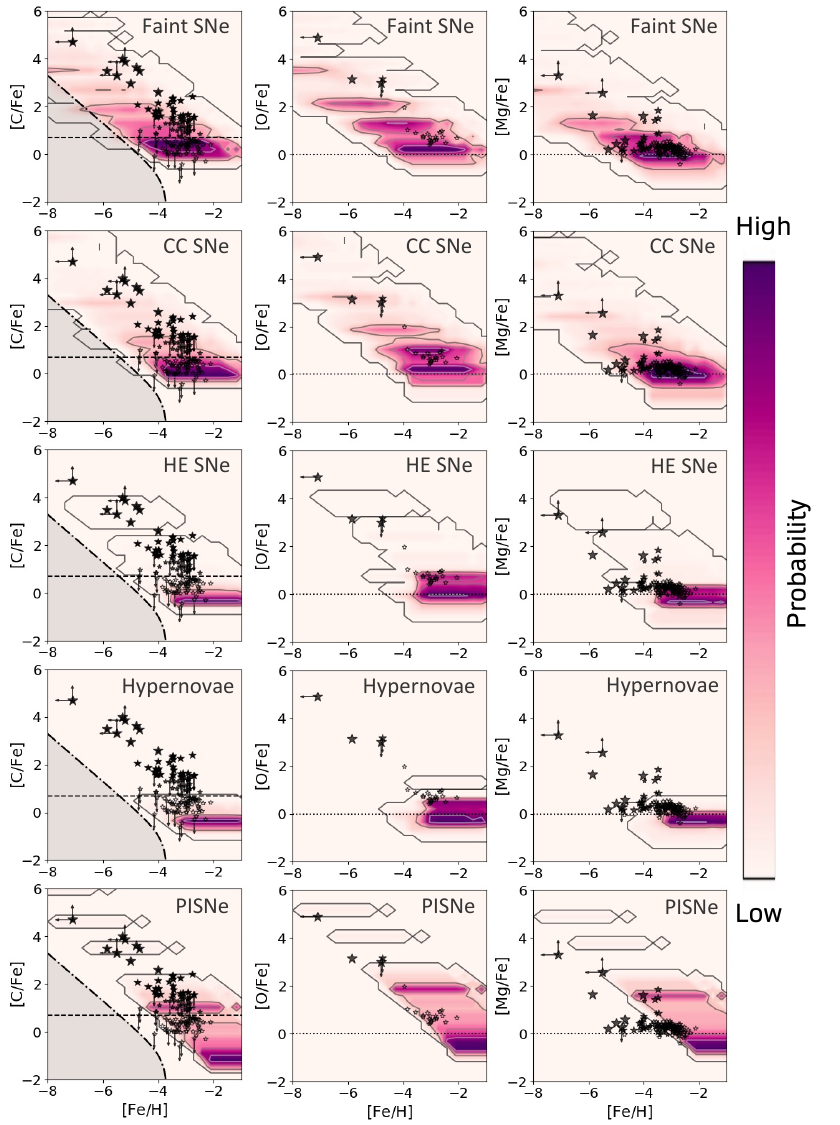}
    \caption{Density maps of the predicted ISM abundances, [X/Fe] vs. [Fe/H], after the explosion of the first generation of stars. Columns are the key chemical elements C, O and Mg; while rows are different Pop~III explosions: faint SN, CC SN, HE SN, hypernovae, and PISN. Star symbols are observed chemical abundances of CEMP-no (filled, the sizes are proportional to the [C/Fe] values) and C-normal (open) halo star. The dash-dotted line in the [C/Fe] diagrams is at $Z_{\rm ISM}=Z_{\rm cr}$. Abundance ratios for the other relevant chemical elements are in Figs.~\ref{fig:PopIII_delta_2} (Si, Ca, Zn) and \ref{fig:PopIII_delta_3} (Al, Ti, Mn). 
    }
    \label{fig:PopIII_delta_1}
\end{figure*}

\begin{figure*}
    \centering
    \includegraphics[width=0.90\textwidth]{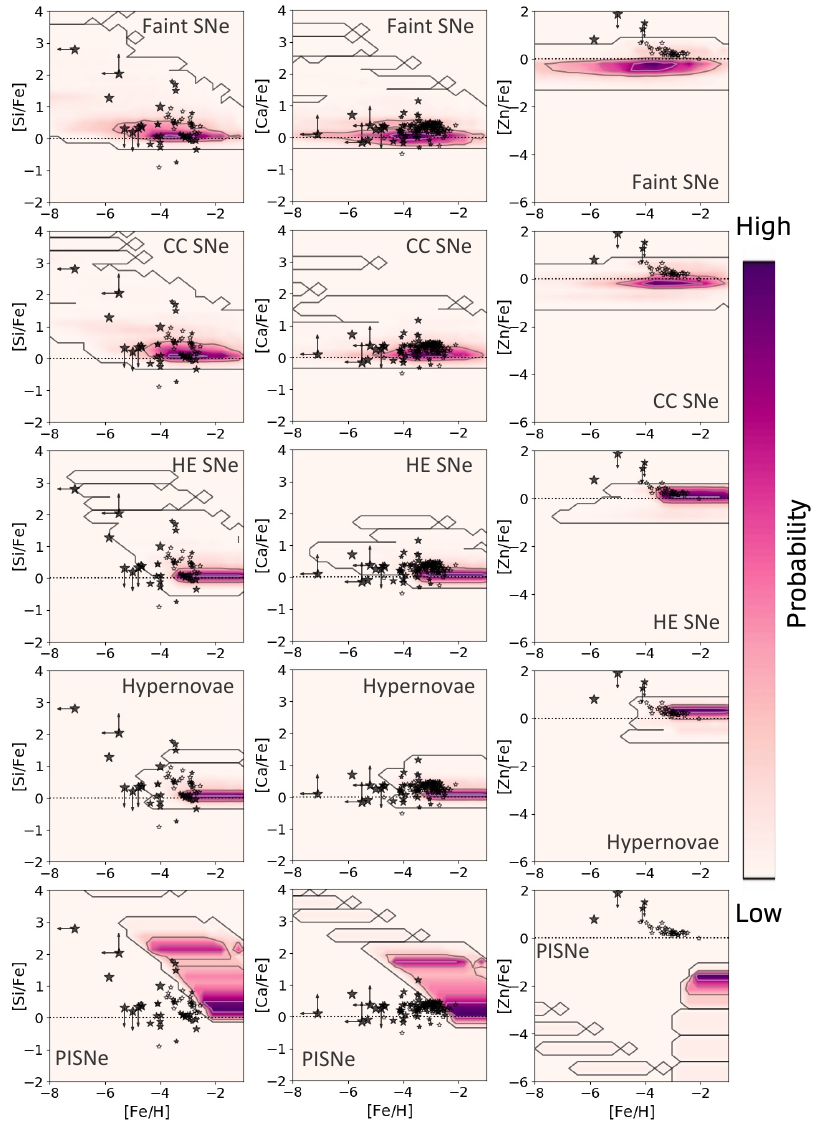}
    \caption{The same as Fig.~\ref{fig:PopIII_delta_1}, but for Si, Ca and Zn. Star symbols are observed chemical abundances of CEMP-no (filled, the sizes are proportional to the [C/Fe] values) and C-normal (open) halo star. Abundance ratios for the other elements are in Figs.~\ref{fig:PopIII_delta_1} (C, O, Mg) and \ref{fig:PopIII_delta_3} (Al, Ti, Mn). }
    \label{fig:PopIII_delta_2}
\end{figure*}

\begin{figure*}
    \centering
    \includegraphics[width=0.75\textwidth]{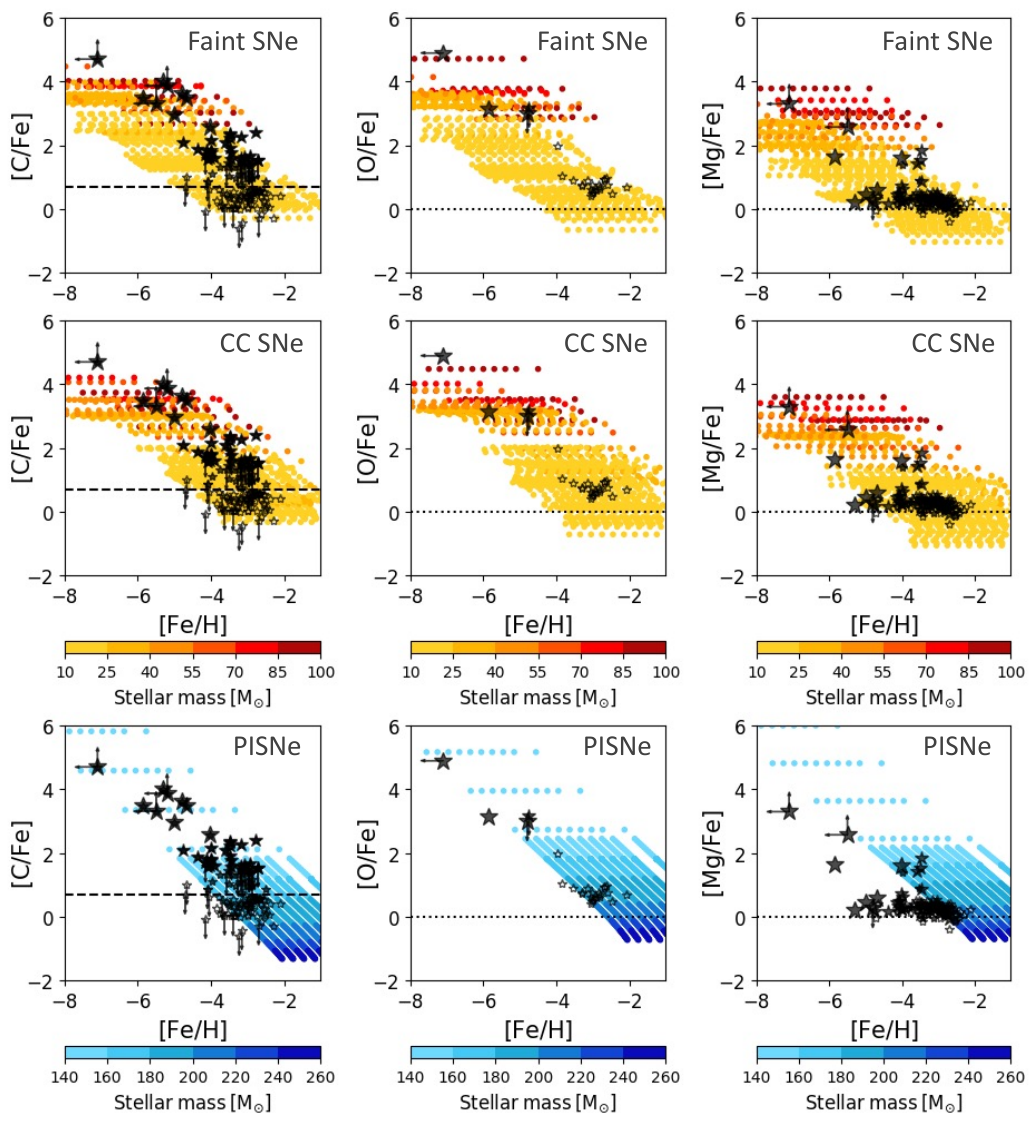}
    \caption{Predicted ISM abundances, [X/Fe] vs. [Fe/H], after the explosion of the first generation of stars with fixed \reply{$f_{\rm mix}=0.063$}. Columns are the key chemical elements C, O and Mg; while rows are different Pop~III explosions: faint SN, CC SN, and PISN. Colors represent the initial stellar masses of the Pop~III progenitor. Star symbols are observed chemical abundances of CEMP-no (filled, the sizes are proportional to the [C/Fe] values) and C-normal (open) halo star. Other elements are shown in Figs.~\ref{fig:PopIII_masses_2} (Al, Si, Ca) and \ref{fig:PopIII_masses_3} (Ti, Mn, Zn).}
    \label{fig:PopIII_masses_1}
\end{figure*}

We start by inspecting the chemical abundances of an ISM solely imprinted by Pop~III stars, i.e. the natal environment of the so-called {\it second generation stars}. 
In Figs.~\ref{fig:PopIII_delta_1} and~\ref{fig:PopIII_delta_2}, we show the density maps for their chemical abundance ratios of C, O, Mg, Si, Ca, and Zn with respect to iron, as a function of iron abundance, [Fe/H]. Additional abundance ratios \reply{of Al, Ti, Mn} are shown in Fig.~\ref{fig:PopIII_delta_3} in Sect.~\ref{sec:appendix_1}. The elements C, O, Mg, Si and Ca, are particularly informative, because they are sensitive to the mass and explosion energy of Pop~III stars. On the other hand, Zn, is a peculiar element for both hypernovae (high values) and PISN (low values). \reply{Conversely, the measured Al is strongly affected by non-LTE effects (see Sect.~\ref{sec:nonLTE}), and Ti is well known to be underestimated by the models of galactic chemical evolution, as it is very sensitive to 3D effects in the explosion of SNe \citep[see][]{Kobayashi2006}. Furthermore, there are no definite measurements available for Mn in CEMP-no stars with $\rm[Fe/H]<-4$ (see Fig.~\ref{fig:VMP_mean}). We completely exclude N from our discussion, because the yields strongly depend on the rotation assumed in the stellar evolutionary model which is highly uncertain. Other elements do not show very strong differences between models, making them less useful than those shown here.}
 
In the first four rows of Figs.~\ref{fig:PopIII_delta_1}, and \ref{fig:PopIII_delta_2}, we show all the possible chemical abundances of an ISM imprinted by Pop~III stars in the mass range [10;100]$\,\mathrm M_{\odot}$ (faint SN, CC SN, HE SN, and hypernovae) and in the bottom row by PISN, with [140; 260]$\,\mathrm M_{\odot}$ and $E_{\rm SN} \in [10^{52}; 10^{53}]$\,erg, see Fig.~\ref{fig:energies}. Note that the iron abundance depends on the free parameters $f_*/f_{\rm dil}$, which are varied in all the plausible parameter space (Sect.~\ref{sec:original}). Conversely, $\rm[X/Fe]=[X/H]-[Fe/H]$, does not depend on them (see eq.~\ref{XsuHiii}) but only on the Pop~III yields, $Y_{\rm X}^{\rm III}$, which clearly depend on both the explosion energy of the Pop~III SNe and the mass of the progenitor star. In the first column of Fig.~\ref{fig:PopIII_delta_1} we show the region in which the ISM metallicity is smaller than the critical value (shaded area). The dash-dotted line indicates where the ISM metallicity equals the critical value on the plane [C/Fe]-[Fe/H]. In this case, $ Z_{\rm ISM}$ is computed considering the amounts of C and Fe only, thus obtaining a lower limit for its value. This ensures that an ISM with [C/Fe] value above the line has a metallicity higher than the critical one, without depending on the assumption for abundances of the other metals.

By inspecting Figs.~\ref{fig:PopIII_delta_1} and \ref{fig:PopIII_delta_2}, we notice that high-energy SNe and hypernovae (row 3 and 4), pollute the ISM with large quantities of Fe. The abundances of their descendants are, therefore, peaked at relatively high $\rm[Fe/H]> -4$ and [C/Fe] $\in (-1, 0)$. These descendants have very small (but non-zero) chances for being C-enhanced. Thus, high-energy SNe and hypernovae alone cannot reproduce all the chemical abundances observed in metal-poor halo stars.
  
On the other hand, the second generation stars formed from the ejecta of faint SNe, CC SNe, and PISNe span $\rm -8\lesssim[Fe/H]\lesssim -1$, and can, therefore, reproduce the iron abundances of all the literature halo sample. Furthermore, they cover a wide range of $\rm -1 \lesssim[C/Fe] \lesssim +5$. The peaks of the [Fe/H] and [C/Fe] distributions are, however, located at different values for different progenitors. The faint SNe descendants show a prominent peak at [Fe/H] $\approx -7$ and [C/Fe] $\approx +3.5$; the CC SNe descendants are more equally distributed in the whole [Fe/H] and [C/Fe] range, showing different peaks at both low and high [C/Fe]; while PISNe descendants have the strongest peak at [Fe/H] $> -2$ and [C/Fe] $< 0$.

To understand the progenitors of CEMP stars, in Fig.~\ref{fig:PopIII_masses_1} we show the predicted [C/Fe], [O/Fe] and [Mg/Fe] values with respect to [Fe/H], color-coded with the initial mass of the Pop~III progenitor star, for one selected mixing efficiency, $f_{\rm mix}=0.063$. 
The [C/Fe] values of the second generation stars strongly depend on the mass of the Pop~III progenitor. For a fixed $f_{\rm mix}$, when the progenitor star explodes as a faint or CC SN, the descendants of the {\it most massive} Pop~III stars show the {\it highest values} of [C, O, Mg/Fe]. If, on the other hand, it explodes as a PISN, this trend is the opposite, the more massive is the progenitor, the lower are the [C, O, Mg/Fe] values of the descendants. Yet, if we vary the mixing efficiency in the range, e.g. $f_{\rm mix} \in [0.039, 0.158]$, this relation between the progenitor masses and the [C/Fe] values of the descendants is not straightforward. Indeed, depending on the mixing efficiency, we find that descendants of progenitors with different masses can have the same [C/Fe] values (see also Fig. \ref{fig:mixing}).

From these figures, we infer that the metal-poor Milky Way halo stars with $\rm[C/Fe]>+2.5$ agree with the chemical abundances predicted for the descendants of Pop~III low-energy SNe, which are also predicted to imprint the ISM with an over-abundance of other light elements: [O/Fe]$>+2$, [Mg/Fe]$>+1.8$, [Si/Fe]$>+1.8$. The carbon abundances of these CEMP-no halo stars also agree with the descendants of the least massive PISN, $m_\star=140M_{\odot}$. However, this doesn't hold for the abundances of the other chemical elements, see e.g. Mg, Ca and Si in Figs.~\ref{fig:PopIII_delta_1} and~\ref{fig:PopIII_delta_2}, thus excluding the possibility for these highly C-enhanced halo stars to be direct descendants of PISN. We can thus conclude that the metal-poor halo stars with [C/Fe] $> +2.5$ are likely the true descendants of Pop~III stars which exploded as low-energy SNe.

At $\rm[C/Fe]< +2.5$, the abundances of the metal-poor halo stars are also consistent with being the descendants of Pop~III stars exploding as more energetic SNe ($10-100\,\mathrm M_{\odot}$), with $E_{\rm SN}$ up to $10^{52}$\,erg, but not of PISNe. However, determining the progenitors of moderately C-enhanced and C-normal metal-poor halo stars is complicated by possible contribution of Pop~II stars (see the following Section). From the first column of Fig.~\ref{fig:PopIII_delta_1} it is evident that, independent of the explosion energy (and progenitor mass), the metals yielded by individual Pop~III SNe typically enable the ISM to reach $\rm {Z_{ISM}\geq Z_{cr}}$, which implies that long-lived Pop~II second-generation stars can form in these environments but also that massive Pop~II stars can start contributing to the ISM enrichment. Furthermore, none of the observed halo stars have $\rm {Z_\star<Z_{cr}}$ (shaded area in Fig.~\ref{fig:PopIII_delta_1}), confirming the critical metallicity scenario for the transition between Pop~III and Pop~II star formation. 

\subsection{Birth environments of Pop~III descendants}

Cosmological models and simulations show that true second generation stars are expected to be rare \citep[see e.g.][]{DeBennassuti2017, Hartwig2018, Liu2021, Rossi2023,Koutsouridou2023}. Hence, here we investigate how the predicted abundance patterns of Pop~III descendants change when their birth environments have also been enriched up to 50\% by normal Pop~II stars, i.e for $f_{\rm PopIII} \geq 0.5$.

\subsubsection{The [C/Fe] ratio}

\begin{figure*}
    \centering
    \includegraphics[width=0.90\textwidth]{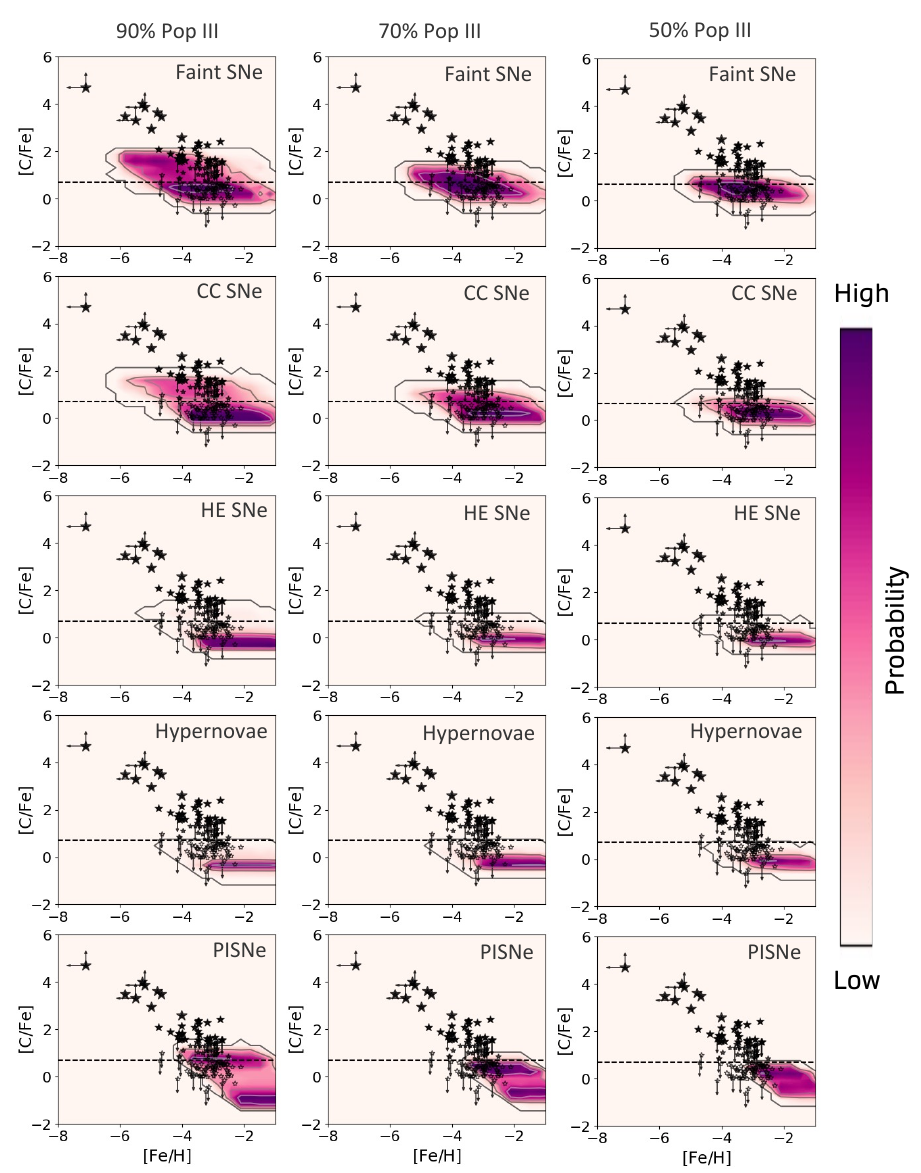}
    \caption{Density map of the predicted ISM [C/Fe] abundance ratios as a function of [Fe/H] with different $f_{\rm PopIII}$: 90\% (left), 70\% (middle), 50\% (right). Explosion energies of Pop~III stars increase from top to bottom. The results obtained with the two sets of Pop~II yields are shown together \citep{Woosley1995,Limongi2018}. Star symbols are observed chemical abundances of CEMP-no (filled, the sizes are proportional to the [C/Fe] values) and C-normal (open) halo star.}
    \label{fig:PopII_carbon}
\end{figure*}

\begin{figure*}
    \centering
    \includegraphics[width=0.9\textwidth]{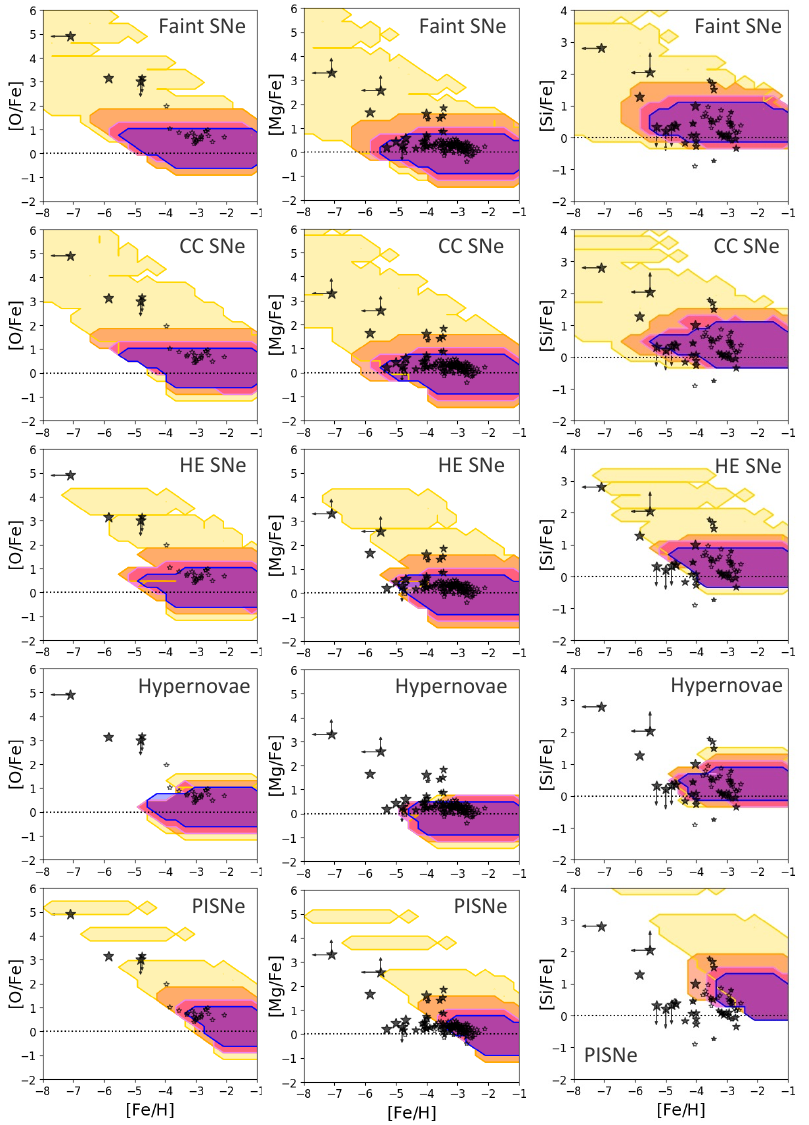}
    \caption{Predicted ISM abundance ratios [X/Fe] vs. [Fe/H], for the elements O, Mg, and Si, after the explosion of Pop~III and Pop~II SNe. Colored areas show different Pop~III contribution to the chemical enrichment: 100\% (yellow), 90\% (orange), 70\% (magenta) and 50\% (purple). The explosion energy of Pop~III stars increases from top to bottom rows. Star symbols are observed chemical abundances of CEMP-no (filled, the sizes are proportional to the [C/Fe] values) and C-normal (open) halo star. Other relevant chemical elements are in Figs.~\ref{fig:PopII_2} (Al, Ca, Zn) and \ref{fig:PopII_3} (Ti, Mn, in Sect.~\ref{sec:appendix_1}).}
    \label{fig:PopII_1}
\end{figure*}

\begin{figure*}
    \centering
    \includegraphics[width=0.9\textwidth]{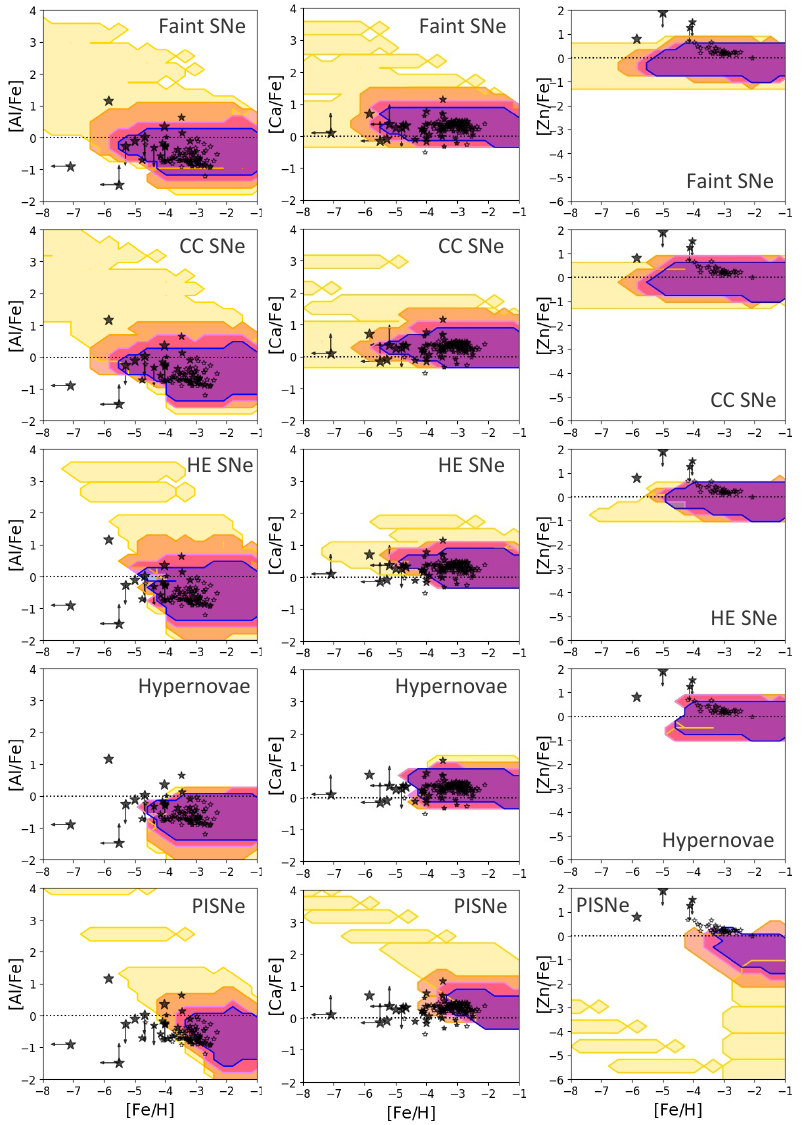}
    \caption{The same as Fig.~\ref{fig:PopII_1} but for Al, Ca and Zn. Star symbols are observed chemical abundances of CEMP-no (filled, the sizes are proportional to the [C/Fe] values) and C-normal (open) halo star. Other relevant chemical elements are shown in Figs.~\ref{fig:PopII_1} (O, Mg, Si) and \ref{fig:PopII_3} (Ti, Mn).}
    \label{fig:PopII_2}
\end{figure*}

In Fig.~\ref{fig:PopII_carbon} we show the [C/Fe] vs [Fe/H] density maps for Pop~III descendants which have also been partially enriched by normal Pop~II stars. We overlap the chemical abundances obtained when using the yields from both \citetalias{Limongi2018} and \citetalias{Woosley1995} since they do not show critical differences.

For increasing Pop~II contribution to the chemical enrichment, [Fe/H] increases while [C/Fe] decreases, moving towards the abundances of the C-normal stars. As shown in Fig.~\ref{fig:PopII_carbon} (first column), a small relative enrichment of 10\% from Pop~II stars is enough to limit the maximum [C/Fe] $\lesssim +2.5$, even in environments predominantly imprinted by low-energy Pop~III SNe. This strongly suggests that halo stars with [C/Fe] $\gtrsim +2.5$ have been enriched {\it only} by low-energy Pop~III stars, and are thus true second generation stars. Indeed, their extreme C-enhancement cannot be reproduced with any contribution ($\geq$10\%) of Pop~II stars, or higher energy Pop~III SN, the only exception being PISN with the lowest mass which however do not produce the other observed abundance ratios (see Sec~\ref{sec:PopII}). 

On the other hand, we can reproduce the halo stars with $\rm[C/Fe]\lesssim +2.5$ with products of low-energy Pop~III SNe and a $\geq$10\% pollution from Pop~II stars. In particular, from Fig.~\ref{fig:PopII_carbon} it is evident that the probability of producing a C-enhanced ISM with the products of low-energy Pop~III SNe decreases as the contribution of Pop~II stars to the chemical pollution increases. We also see that high-energy Pop~III SNe can, with small probability, imprint the ISM up to [C/Fe]$\approx +1.5$ if the contribution of Pop~II stars is $\lesssim 10\%$. The same is true for PISN enrichment but in this case it is limited to [Fe/H] $\geq -4$. 
Finally, we note that gaseous environments predominantly imprinted by Pop~III hypernovae cover a broad iron range, $\rm[Fe/H]\geq -4.5$ but always have $\rm[C/Fe]<+0.7$, regardless of the Pop~II contribution. 

Based on the [C/Fe] abundance ratio, we can only conclude that the most C-enhanced halo stars are true second generation stars. On the other hand, the CEMP-no stars with [C/Fe] $<+2.5$ could be polluted by Pop~III stars exploding with different energies and Pop~II stars at different levels. Thus, we need to investigate the imprint of the first stars also with heavier chemical elements.

\subsubsection{Elements beyond carbon}

In Fig.~\ref{fig:PopII_1} we show the predicted range of [X/Fe] vs [Fe/H] for the $\alpha$-elements O, Mg and Si, with varying Pop~III contribution (from 100\% down to 50\%), and for increasing explosion energy of Pop~III SNe (top to bottom rows). The same is shown in Fig.~\ref{fig:PopII_2} for Al, Ca, and Zn (Fig.~\ref{fig:PopII_3} for Ti, and Mn). In general, the predicted abundance ratios for elements lighter than Ca follow a similar trend to [C/Fe], i.e. the maximum [X/Fe] values decrease as the Pop~II contribution increases, moving towards the abundances of C-normal stars. 

Most of the highly C-enhanced halo stars ($\rm[C/Fe]> +2.5$, largest points in Figs.~\ref{fig:PopII_1}, \ref{fig:PopII_2} and \ref{fig:PopII_3}) also have high values of [O, Mg, Si/Fe] and these enhancements agree only with 100\% enrichment from Pop~III SNe, either faint, core-collapse or high-energy SNe. On the other hand, we predict too high [Al/Fe] values for the highly C-enhanced stars. However, the 3D and non-LTE corrections for [Al/Fe] are estimated to be $\gtrsim +0.6$~dex for these Fe-poor stars (\citealp[see][]{Nordlander2017a}; see Sect. \ref{sec:nonLTE}). At the lowest $\rm[Fe/H]<-5$, the NLTE corrected data could therefore agree with an enrichment by Pop~III faint and core-collapse SNe, while for the descendants of high-energy SNe we predict [Al/Fe] values which cannot agree, even after the correction. 

The highly C-enhanced stars of our literature sample also show high [Zn/Fe] values ($\gtrsim 0.8$), though there is only one star with a real [Zn/Fe] measurement \citep{Ezzeddine2019} while for the other there exist only upper limits. Our predictions for faint, CC and HE SNe descendants are therefore consistent with the [Zn/Fe] values of the highly C-enhanced halo stars. The only star with a finite value for [Zn/Fe] is only marginally consistent with our predictions for different chemical elements. To explain its uncommon abundance pattern, peculiar SN explosion mechanisms, such as the aspherical SN explosion, have been indeed proposed \citep[e.g. see][]{Ezzeddine2019}.

To conclude, the abundance ratios of the descendants shown in Figs.~\ref{fig:PopII_1}, \ref{fig:PopII_2} and \ref{fig:PopII_3}, confirm that the most C-enhanced and Fe-poor halo stars have been most likely imprinted by a single or few Pop~III low-energy SNe ($E_{\rm SN} < 3 \times 10^{51}$erg). 

The C-enhanced stars with $\rm [C/Fe] \leq +2.5$ have [O, Mg, Si, Al, Ca/Fe] values in agreement with the descendants of Pop~III SNe, either faint, core-collapse or high-energy, with a contribution from Pop~III stars down to 70\%, for the ones with $\rm [C/Fe] > +1.5$, and down to 50\%, for the ones with $\rm [C/Fe] \leq +1.5$. 
Conversely, the range of abundance ratios predicted for the descendants of hypernovae and PISNe do not match the observed one. As before, the high values of [Zn/Fe] for the CEMP stars with [C/Fe] $\leq +2.5$ is mainly based on upper limits and, therefore, is in agreement with all our predictions for Pop~III descendants, but only marginally with PISNe.

Finally, the C-normal stars of our literature sample agree with all the abundance ratios predicted by our model (Figs.~\ref{fig:PopII_1}, \ref{fig:PopII_2} and \ref{fig:PopII_3})\footnote{With the exception of [Ti/Fe] about which we will discuss in the next Section.} for the descendants of Pop~III stars with a substantial, $\leq 50\%$, contribution from Pop~II stars. Only the descendants of PISNe are not able to reproduce the abundances of C-normal stars.

To conclude, while the progenitors of the most C-enhanced stars are likely single or few massive primordial SNe, the abundances of moderately C-enhanced ([C/Fe] $\lesssim +2.5$) and C-normal stars are consistent with both the enrichment from primordial Pop~III SNe and/or from a subsequent generation of Pop~II stars.

\subsection{The complete abundance pattern}

\label{sec:abu_pattern}

\begin{figure*}
    \centering
    \includegraphics[width=\linewidth]{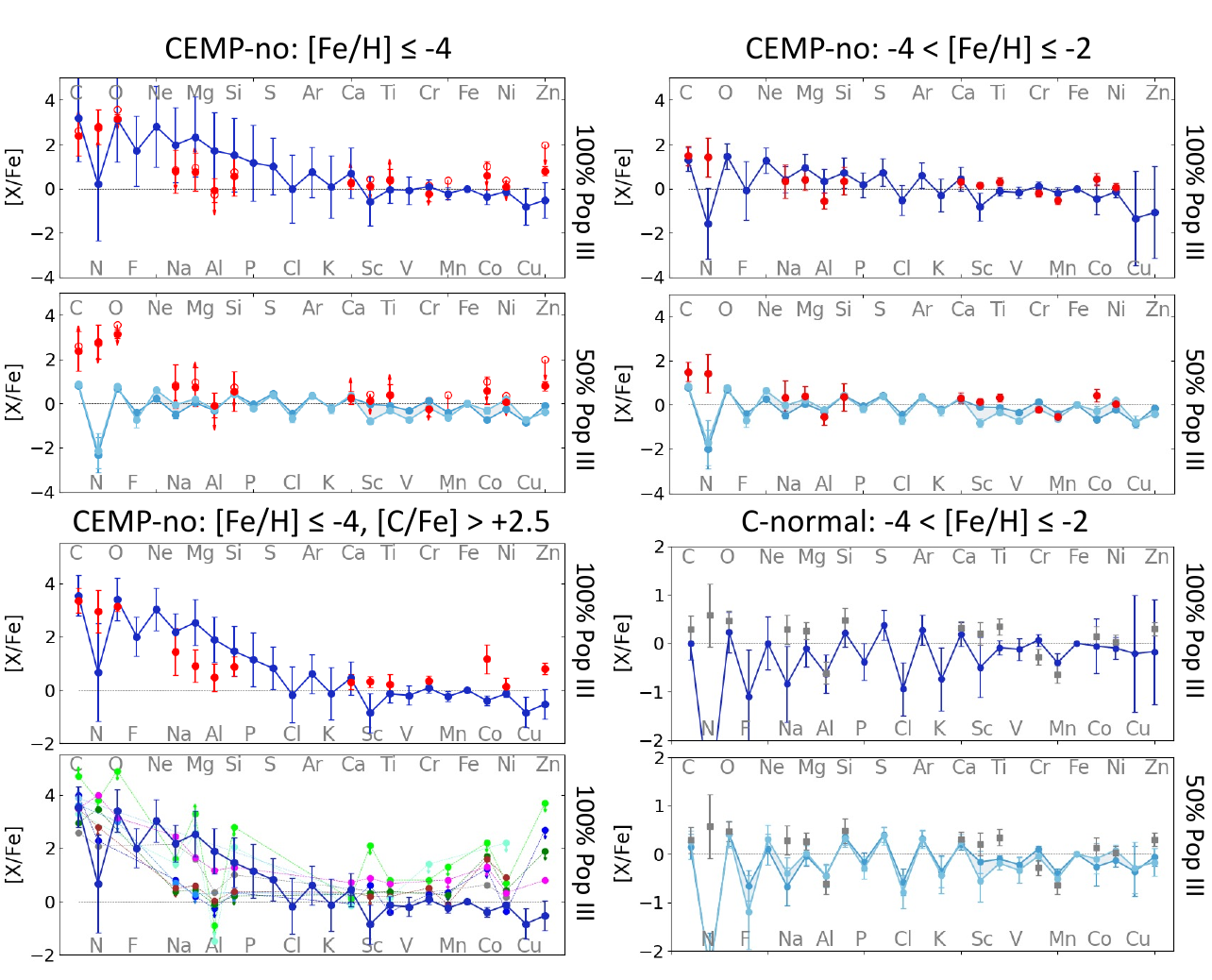}
    \caption{\reply{Mean chemical abundance patterns for our models (blue) with 100\% and 50\% contribution from Pop~III stars that predict: CEMP-no descendants with $\rm-7.5\le [Fe/H]\le -4$ (top left); CEMP-no descendants with $\rm-4<[Fe/H]\le-2$ (top right); C-normal descendants with $\rm-4<[Fe/H]\le-2$ (bottom right); compared with the mean measured abundances of CEMP-no stars (red) and C-normal stars (grey; see Sect. \ref{data} and Fig.~\ref{fig:VMP_mean}). \emph{Bottom left panels}: Mean chemical abundance patterns predicted for second generation stars (blue, 100\% Pop~III pollution) with $\rm-7.5\le [Fe/H]\le -4$ and $\rm+2.5\leq [C/Fe]\leq +5.5$, compared with the mean measured abundances of extremely C-enhanced stars, excluding the upper/lower limits (red, upper panel) and the abundances of individual stars (lower panel, colors as in Fig.~\ref{fig:CEMPeCno}). Blue error bars are the standard deviation between models. For $f_{\rm PopIII} = 50\%$ the abundance patterns are calculated using both PopII yields from \citetalias{Woosley1995} (darker blue) and \citetalias{Limongi2018} (lighter blue).
    The coloured area between them represents an intrinsic uncertainty due to the choice of the Pop~II yields.}}
    \label{fig:abu_pattern_tot}
\end{figure*}

In the previous Sections we investigated how a single Pop~III star pollutes with its chemical products the ISM in which it explodes and how the resulting chemical abundances change with the Pop~II contribution. Our results show that C-normal environments (or stars) can be either imprinted by single Pop~III SN or predominantly polluted by normal Pop~II stars. Here we aim at discriminating among these two possibilities by exploiting all the different chemical elements measured.

In Fig.~\ref{fig:abu_pattern_tot}, we show the mean chemical abundance patterns of all predicted Pop~III 100\% and 50\% descendant stars, distinguishing between the CEMP-no ones with [Fe/H] $\in [-7.5; -4]$ (top left) and $\in (-4; -2]$ (top right) and the C-normal ones with [Fe/H] $\in (-4; -2]$ (bottom right), \reply{compared with the average abundances of observed stars}. \reply{We also show the average abundance pattern of second generation descendants which have [Fe/H] $\in [-7.5; -4]$ and [C/Fe] $\in[+2.5; +5.5]$ (bottom left), compared with the average abundances of the observed stars (upper) and the abundances of each single star (lower).}
Computing the mean chemical abundances of the mini-halos, we are assuming that the five types of primordial SNe \reply{and the four mixing efficiencies}, already discussed, are equiprobable. \reply{Basically, we selected all the models that produce the considered chemical abundances and averaged over their abundance ratios. For instance, to produce the abundances depicted in the top right section of Fig.~\ref{fig:abu_pattern_tot} we averaged over all the models that produce $\rm[Fe/H]\in (-4; -2]$ and $\rm[C/Fe]>+0.7$.}

For the CEMP-no stars in the top left of Fig.~\ref{fig:abu_pattern_tot}, the predicted abundance ratios, [X/Fe], of elements lighter than Ca decrease with increasing Pop~II contribution.
The average [Cu/Fe] and [Zn/Fe] increase as the contribution from Pop~II stars increase.
This effect is mainly due to the fact that the most copper- and zinc-deficient descendants, which are the ones enriched by primordial PISN, become C-normal when we add the contribution from Pop~II stars and do not contribute anymore to the average abundance pattern of CEMP-no descendants.
On the contrary, the abundances of C-normal descendant stars (bottom panels of Fig.~\ref{fig:abu_pattern_tot}) do not change significantly with an increasing Pop~II contribution. 
In all the cases depicted in Fig.~\ref{fig:abu_pattern_tot}, the standard deviation of the predicted chemical abundances is significantly reduced when the Pop~II contribution increases. 

Our models always underestimate the [N/Fe] ratio. This is partly due to the fact that PISNe ejecta present a strong odd-even effect, always producing $\rm[N/Fe]\le 0$ in the ISM \citep[see][]{Heger2002a,Salvadori2019}, and partly due to the difficulty of modeling the nucleosynthesis of N in the stars with $\rm M \in [10; 100] M_{\odot}$. Indeed, the amount of N synthesized in the stars strongly depends on the mixing between the internal layers, which is usually achieved with stellar rotation (see \citealt{Iwamoto2005a,Kobayashi2006,Heger2010} and \citealt{Chiappini2005,Limongi2018} for a comparison between the yields for rotating and non-rotating stars). The yields adopted in this work are for non-rotating stars and predominantly present $\rm [N/Fe] < 0$.

At the lowest $\rm[Fe/H]\le-4$, the measured abundances of CEMP-no stars are only consistent with theoretical predictions of 100\% Pop~III contribution, as we pointed out in the previous Sections. CEMP-no stars with $\rm-4<[Fe/H]<-2$, on the other hand, have chemical abundances that are consistent with either being second generation stars, or having Pop~III enrichment at a $\gtrsim70\%$ level, with a partial Pop~II pollution. Our model overestimates the [Mg/Fe] and [Al/Fe] ratios, relative to observations, but when non-LTE corrections are applied, this discrepancy is expected to disappear (see Sect.~\ref{sec:nonLTE}).
The predicted [Zn/Fe] average is lower than the observed one. As discussed in the previous Section, the yields of normal SNe are not able to reproduce the observed high [Zn/Fe] values and, moreover, the abundances of PISNe descendants, which have $\rm [Zn/Fe] \ll 0$, strongly lower the average [Zn/Fe]. However, the average observed [Zn/Fe] is an upper limit and there is only one finite measurement, which is just at the edge of the values predicted by our model (Fig.\ref{fig:PopII_2}).

\reply{The abundances depicted in the bottom left panels of Fig.~\ref{fig:abu_pattern_tot} are a subclass of the abundances in the top left panels. The measured abundances include only the extremely C-enhanced stars ([C/Fe] $>+2.5$) which we predict to be the direct descendants of Pop~III stars. First of all, we note that the measured abundances are very different from star to star, most of all for C, Mg, Al, Si and Zn. Moreover, a great number of them is with upper or lower limits: for instance, we have no definite measurement for [Mn/Fe] and just one for [Zn/Fe]. Therefore, the average abundances for [C/Fe] $>+2.5$ shown in the upper panel might not be representative of the real values of [X/Fe].} 
\reply{In this case we show only the 100\% Pop~III SNe case because with a contribution of Pop~II stars $\geq 10\%$ the descendants reach a maximum [C/Fe] $<+2.5$. The average abundance ratios predicted for C, O, Na, Ca, Cr and Ni are consistent with the measured ones. Our model predicts higher [Mg/Fe] and [Al/Fe] with respect to the observations, but we expect this discrepancy to be relieved if non-LTE correction are applied (see Sect.~\ref{sec:nonLTE}). Conversely, our model can't reproduce the observed abundances of [Sc/Fe], [Co/Fe] and [Zn/Fe] \citep[see also][]{Kobayashi2020a}.}

The measured chemical abundances of the C-normal stars in the range [Fe/H] $\in (-4; -2]$, on the other hand, agree with our predictions for the descendants enriched by Pop~III stars at a $\gtrsim50\%$ level. 
However, if the birth environments of C-normal descendants are predominantly ($>$ 50\%) polluted by Pop~III stars, the predicted scatter is higher than what observed in C-normal stars. Furthermore, the agreement between the average and the observed abundances is better for some elements, C, O, Sc, Co and Zn, with a contribution of Pop~II stars at the $\sim50$\% level. For C-normal descendants we predict a lower [Na/Fe] average, but the non-LTE corrections (see Sect. \ref{sec:nonLTE}) should lower the observed abundances. Finally, we predict a smaller [Ti/Fe] with respect to the observed one. Ti is lightly affected by non-LTE effects and in general the stellar evolution models underestimate it \citep[see][ and references therein]{Kobayashi2006}.

\subsection{The star-to-star scatter}

\label{sec:scatter}

\begin{figure*}
    \includegraphics[width=\textwidth]{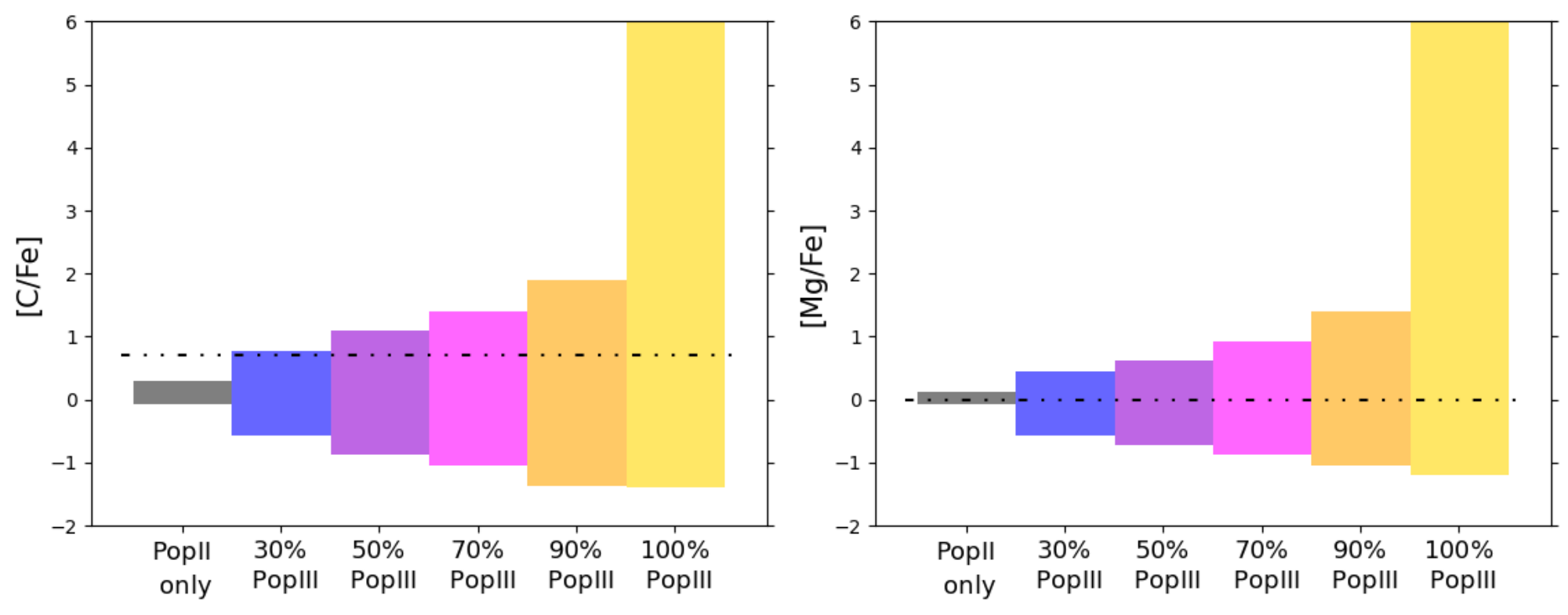}
    \caption{Maximum extent of the abundance ratios [C/Fe] (left) and [Mg/Fe] (right) for Pop~III descendant stars as a function of relative Pop~III pollution. reply{The mixing efficiency is fixed, $f_{\rm mix}=0.063$.} Coloured areas represent different levels of enrichment from Pop~III stars, $f_{\rm PopIII}$: 100\% (yellow), 90\% (orange), 70\% (magenta), 50\% (purple), 30\% (indigo), and only Pop~II stars (grey). The dash-dotted lines correspond to [C/Fe] $=+0.7$ (left) and [Mg/Fe] $=0$ (right). }
    \label{fig:scatter}
\end{figure*}

The last and most conclusive result of our work concerns how the maximum scatter in abundance ratios is dependent on the relative pollution from Pop~III stars, $f_{\rm PopIII}$. In Fig.~\ref{fig:scatter} we show the predicted dispersion of [C/Fe] and [Mg/Fe], with respect to $f_{\rm PopIII}$, with a fixed mixing efficiency for Pop~III stars of $f_{mix}=0.063$. The pure Pop~III descendants, $f_{\rm PopIII}=100\%$, show a dispersion $> 5$ dex in the abundance ratios [C/Fe] and [Mg/Fe], as well as in the other chemical elements lighter than Ca. These abundance ratios are very dependent on \reply{the mass and SN explosion energy of the progenitor} and therefore vary over a wide range of values. \reply{We point out that the predicted dispersion for the descendants of Pop~III stars does not change if we use all the four mixing efficiencies together. This means that the scatter is driven by the different initial masses and SN explosion energies of Pop~III stars.} As Pop~II stars contribute more to the pollution of the ISM, they wash out the diverse chemical peculiarities of the different primordial progenitors and the dispersion between different descendants is reduced. Finally, when the contribution from Pop~III stars is negligible (Pop~II only case), the abundances of the descendants almost correspond to the solar values. 

To conclude, with our model, we predict that the scatter in [C/Fe] and [Mg/Fe] ratios is maximum for Pop~III only enriched environments and that it decreases as the contamination from Pop~II stars increases. 

The \emph{scatter diagnostic} can also be applied to high-redshift absorption systems for which the measurement of the hydrogen column density is not possible, because this prediction only uses abundance ratios between different metals. It allows, without the classical comparison with [Fe/H], to understand if an absorption system has Pop~III fingerprints in its gas (see Sodini et al. 2023, in prep.).

\section{Discussion}

\label{discussion}

\subsection{Model's generality and comparison with previous works}

\begin{figure*}
    \centering
    \includegraphics[width=\linewidth]{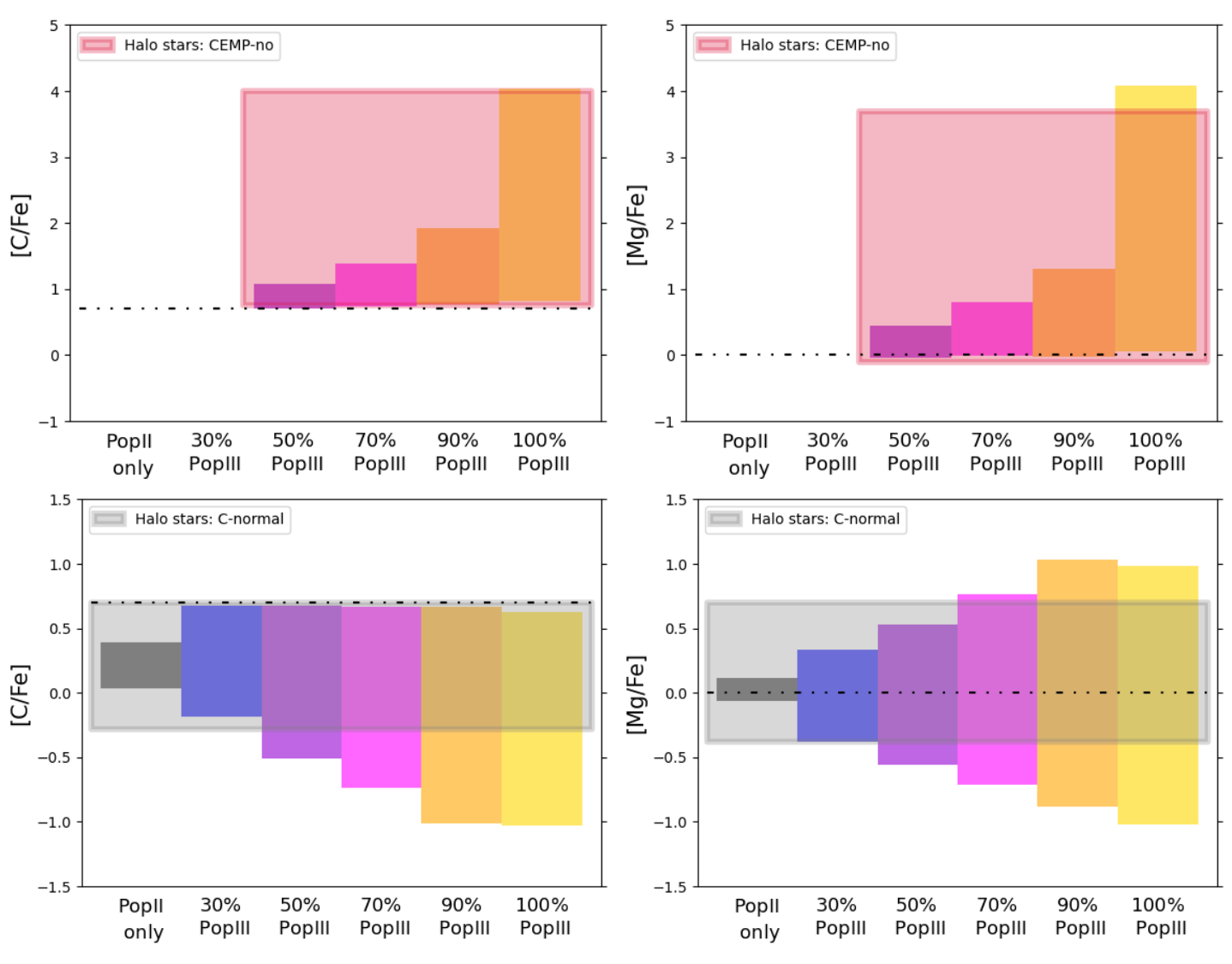}
    \caption{Maximum extent of the abundance ratios [C/Fe] (left) and [Mg/Fe] (right) for Pop~III descendants as a function of relative Pop~III pollution. \reply{The mixing efficiency is fixed, $f_{\rm mix}=0.063$.} Coloured areas represent different levels of enrichment from Pop~III stars: 100\% (yellow), 90\% (orange), 70\% (magenta), 50\% (purple), 30\% (indigo), and only Pop~II stars (grey). Overlapping shaded areas are the maximum scatter between the measured abundances of the metal-poor halo stars, divided in two categories: CEMP-no stars (red) and C-normal stars (grey) from the literature sample, without considering the upper/lower limit measurements (Sect.~\ref{data}). The dash-dotted lines correspond to [C/Fe] = +0.7 (left) and [Mg/Fe] = 0 (right).
    }
    \label{fig:scatter_yong}
\end{figure*}

The parametric model proposed in this paper is very general, which makes it suitable for applications on a broad range of topics related to early chemical enrichment. 
Indeed it can interpret the chemical fingerprints left by the first SN explosions in both long-living stellar descendants \citep[e.g.][]{Skuladottir2023} and in more distant gas clouds, which can be observed as absorption systems \citep[e.g.][Sodini et al. in prep.]{Salvadori2023}. Following the results of cosmological simulations of primordial star formation \citep[e.g.][]{Hirano2014,Susa2014,Hirano2017}, our model assumes that a single Pop~III star forms in each proto-galaxy. This simple but physically motivated hypothesis allows us to understand how the chemical abundances of the Pop~III star descendants vary with Pop~III stellar properties: their initial mass, the mixing efficiency (see Sect.~\ref{sec:appendix_2}) and the SN explosion energy. For the first time, we demonstrated the importance (and degeneracy) of these three unknowns to interpret the entire chemical abundance patterns of ancient stars. This is the key to interpret the results of more sophisticated semi-analytical models which assume different Pop~III IMF and energy distribution functions to follow the formation and evolution of different galaxies \citep[e.g.][]{Rossi2023,Koutsouridou2023}.

Our findings for an ISM solely enriched by Pop~III SNe are in excellent agreement with the results of \citet{Cooke2014} and \citet{Welsh2020}, who studied the abundance of some specific chemical elements (C, Mg, Ca, Ni, Fe) after the pollution of Pop~III stars only. However, here we also investigate how the contamination of normal Pop~II stars can affect the abundance pattern of the Pop~III enriched environments, comparing \emph{all} the chemical abundances measured for the halo stars with the ones predicted by our model.
In particular, we show that all stars with [C/Fe] $\geq +2.5$ are genuine second-generation stars. Our results show that the probability of a gaseous environment (star) for being also imprinted by Pop~II stars increases as the [C/Fe] (and other abundance ratios as [Mg/Fe]) decreases. In other words, the peculiar and variegate abundance pattern left in the ISM by Pop~III SNe is gradually washed out by the dominant pollution from different generations of normal Pop~II stars, which shrink the abundance ratios. Thus, we suggest that C-normal metal-poor halo stars might be the result of this dominant Pop~II contribution, which is consistent with the results of both the metal-enrichment model developed by \citet{Liu2021} and cosmological semi-analytical models for the Milky Way formation \citep[][]{DeBennassuti2017,Koutsouridou2023}. 

But can we really be sure that C-normal stars are not truly second-generation objects? As we pointed out in Sect.~\ref{sec:PopII}, our model predicts that among C-normal (and [C/Fe] poor) stars there might be some second-generation stars, solely imprinted by Pop~III SNe. This result is in line with the one of \citet{Welsh2020}, who interpret the origin of C-normal stars with $\rm [Fe/H]<-2.5$ using a stochastic model for Pop~III chemical enrichment. However these authors, by showing that their [Mg/Ca] and [Ni/Fe] are well fitted by multiple Pop~III high-energy SNe, concluded that C-normal stars are all second generation stars. 
Conversely, our analysis of the entire chemical abundance pattern (15 elements in total) seems to suggest that this is not the case, since the star-to-star scatter should have been much larger in the case of a pollution driven by Pop~III SNe only. 

In Fig.~\ref{fig:scatter_yong} we compare the observed star-to-star scatter with the predicted one for [C/Fe] and [Mg/Fe], separating the CEMP-no (top panels) and the C-normal stars (bottom panels). The shaded areas represent the star-to-star scatter of the stars in our literature sample. The theoretical scatter is computed separately for CEMP-no and C-normal descendants, \reply{fixing the mixing efficiency $f_{\rm mix}=0.063$}, by randomly selecting the same number of descendants in our model as available in the literature. We repeated this random procedure 100 times and averaged between the 100 minimum and maximum [X/Fe] values. In Fig.~\ref{fig:scatter_yong}, we see that the star-to-star scatter of CEMP-no stars is consistent with the one predicted for the birth environments of second-generation stars. On the other hand, the star-to-star scatter of C-normal stars  is consistent with the dispersion predicted for environments imprinted by Pop~III stars at a $\le 50\%$ level.

\begin{figure}
    \centering
    \includegraphics[width=\linewidth]{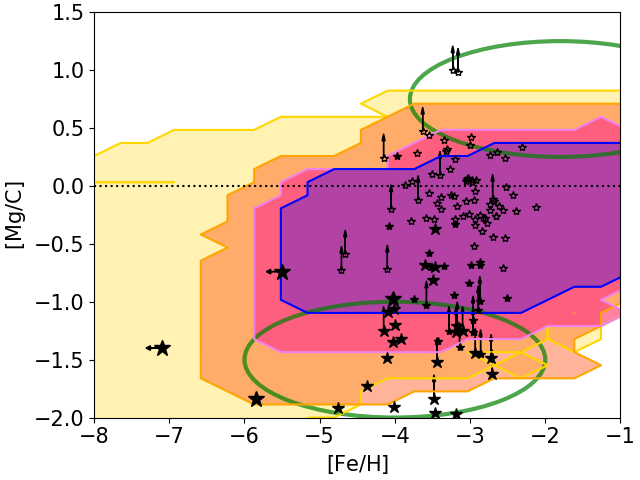}
    \caption{Predicted abundance ratios [Mg/C] vs. [Fe/H] for Pop~III descendants, compared with the literature sample. Colored areas show different Pop~III contribution: 100\% (yellow), 90\% (orange), 70\% (magenta), 50\% (purple). Star symbols are observed chemical abundances of CEMP-no (filled, sizes are proportional to their [C/Fe]) and C-normal (open) halo star. Green ellipses show the Pop~III mono-enriched areas identified in Fig.~11 of \citet{Hartwig2018}.}
    \label{fig:hartwig}
\end{figure}

Recently, \citet{Hartwig2018} proposed a new diagnostic to identify stars (or ISM) mono-enriched by a single Pop~III SN. They show that in the [Mg/C] vs [Fe/H] diagram, stars with [Mg/C] 
$\approx -1.5$ and $\approx 0.7$ have the highest probability to be mono-enriched \citep[see Fig.~11 and~15 in][]{Hartwig2018}. Do we find the same results for our environments enriched by a single Pop~III star? In Fig.~\ref{fig:hartwig} we compare the predictions of \citet{Hartwig2018} with those of our model for the birth environments of second generation 
mono-enriched stars and with other Pop~III descendants ($50-90\%$ Pop~III polluted). The area populated by Pop~III mono-enriched stars is significantly wider than what was found in \citet{Hartwig2018}. Furthermore, we do not predict second generation stars at $\rm[Mg/C]>+0.75$. These inconsistencies have a double explanation: firstly, \citet{Hartwig2018} only explored the imprint from low-energy (faint and core-collapse) Pop~III SNe with initial masses from 10 to 100~$\mathrm{M_{\odot}}$; 
secondly, they assumed the yields from \citet[][]{Kobayashi2012,Nomoto2013,Ishigaki2014} while we adopt those from \citet{Heger2010}.

We also note that our predictions for an ISM contaminated by Pop~II stars at 
a level between $10\%$ and $50\%$ populate the [Mg/C] diagram in the same areas identified by \citet{Hartwig2018} for Pop~III mono-enriched stars. From Fig.~\ref{fig:hartwig} we see that the region occupied {\it only} by Pop~III mono-enriched stars is very narrow \citep[see also][]{Hansen2020}. We see that only two CEMP-no stars with $\rm[C/Fe] \geq +1.5$ in our entire literature sample are unambiguously 2G mono-enriched stars, based on Fig.~\ref{fig:hartwig}, while all the others are consistent with being either 2G stars or partly contaminated by normal Pop~II stars. 
CEMP-no halo stars in Fig.~\ref{fig:hartwig} that show $\rm[Mg/C]< -1.8$ and $\rm[Fe/H]> -4.5$ might have been imprinted also by Pop~III/Pop~II AGB stars, which are not considered in this work. \reply{Yet, the contribution of Pop~III AGB stars has only a minor effect on the gas previously enriched by a SN because the amounts of carbon ejected by AGB stars is lower than what ejected by SNe in the early universe \citep[][]{Rossi2023}. The minor variations obtained by including Pop~III AGB stars are well within other uncertainties of the model, e.g. in the chemical yields (Vanni et al. in prep).} On the contrary, the two C-normal stars with extremely high $\rm[Mg/C]\gtrsim +1.0$ might have formed after the explosion of the most massive ($m_\star\approx 40 \rm M_{\odot}$) Pop~II stars, which eject high amounts of Mg \citep[e.g. see][]{Salvadori2019}.

\subsection{Can we trust the data?}

\label{sec:nonLTE}

The predicted [C/Fe] and [Fe/H] ratios for the descendants of Pop~III faint and core-collapse SNe only agree with the abundances of the most C-rich CEMP-no halo stars when the parameter $\dfrac{f_*}{f_{\rm dil}}$ is maximum, which means having either a high Pop~III star-formation efficiency and/or metals diluted in a small portion of the primordial gas. However, we haven't corrected the chemical abundances of the observed stars for the non-LTE and 3D effects: ideally this should be done on each singular star, but we can estimate the overall corrections.  

The C and Fe abundances of our literature sample are not corrected for non-LTE or 3D effects, except for a few stars among the most C-enhanced ones: the \cite{Christlieb2004} (only Fe), \cite{Caffau2011a} (both C and Fe), \cite{Keller} (both), \cite{Bonifacio2018} (only C), \cite{Starkenburg2018} (only C) and \cite{Ezzeddine2019} (only Fe) stars. However, non-LTE 3D models of stellar atmospheres have become more and more sophisticated recently and the chemical abundances, even if already corrected, might deserve another revision. The 3D and non-LTE corrections of the iron abundance, if done consistently, are opposite (the 3D is negative and the non-LTE is positive), resulting in a total shift for [Fe/H] of the order of -0.05 to +0.15~dex \citep[see][]{Amarsi2019,Norris2019}. The corrections to the carbon abundance are more severe, most of all for low metallicity stars, and can be up to -1.0~dex \citep[see][]{Amarsi2019,Norris2019}, leading to a total correction to [C/Fe] of the order of -0.5 to -1.0~dex. With these corrections, the CEMP-no halo stars with [C/Fe]$>+2.5$ would still agree with being the descendants of Pop~III low-energy SNe, but the parameters, such as the initial mass of the progenitors, would change.
Nevertheless, this would still exclude the possibility of being primarily enriched by high-energy Pop~III SNe (see Fig.~\ref{fig:PopIII_masses_1}) and Pop~II stars. 

The 3D and non-LTE corrections are also not negligible for the abundances of O, Na, Mg and Al. The [O/Fe] ratios of C-normal stars in our literature sample, at $-4 <$ [Fe/H] $< -2$, are corrected for 3D effects, as in \citet{Cayrel2004a}, of $-0.23$~dex. The O abundances of CEMP-no stars at lower [Fe/H] are not corrected. For them the corrections should be higher with respect to the more Fe-rich ones, up to $-0.6$~dex \citep[see][]{Amarsi2019}. [Na/Fe], [Mg/Fe] and [Al/Fe] are not corrected for the entire literature sample. \citet{Cayrel2004a} estimate a correction for [Na/Fe] of up to $-0.5$~dex, which can improve the agreement with the models for C-normal stars (see Fig.~\ref{fig:abu_pattern_tot}), and for [Mg/Fe] of $\sim +0.27$~dex \citep[see also][]{Andrievski2010}. Such an increase in the [Mg/Fe] ratio would make our models underestimate it for C-normal stars. This might be an indication that these stars have been polluted mostly by massive ($\sim 30\,\mathrm M_{\odot}$) Pop~II stars, which eject high amounts of Mg with respect to Fe (this peculiar feature is washed out by the integrated contribution of the least massive Pop~II stars). Ultimately, the correction to be applied on [Al/Fe] is $\gtrsim +0.6$~dex \citep[see e.g.][]{Andrievski2008,Nordlander2017a}, but it strongly depends on the metallicity of the star, the lower is the metallicity, the higher is the correction.

\section{Conclusions}

\label{conclusion}

The metal-poor stars in the Galactic halo offer a unique opportunity to identify the chemical fingerprints of the first stars and hence understand their properties. The most iron-poor and carbon-rich halo stars are commonly thought to be true second generation stars, i.e.~stars that have been imprinted solely by Pop~III stars. On the other hand, the debate is still open for the more iron-rich halo stars, which are thought to be either imprinted only by Pop~III stars or also by normal Pop~II stars. 

Here we aim at finding the peculiar chemical imprints left by the first stars, and to determine which of the halo stars are real descendants of Pop~III stars, i.e. if $\gtrsim 50\%$ of their metals are produced by Pop~III stars. In order to achieve our objectives, we further improved and extended the parametric model developed by \citet{Salvadori2019}, and explored the chemical abundances in the first star-forming structures after the pollution of: (i)~one Pop~III star; (ii)~also the Pop~II stars which formed subsequently. Comparing the chemical abundances resulting from our model with literature halo stars, we find that:

\begin{itemize}
    \item The most C-enhanced ([C/Fe] $> +2.5$) halo stars have chemical abundances that agree with an imprint from only one primordial Pop~III star exploding with low energy ($< 2 \times 10^{51}$ erg). 
    \\
    \item C-enhanced metal-poor halo stars with $+0.7 <$ [C/Fe] $< +2.5$ are likely born in environments polluted by both Pop~III and Pop~II stars where Pop~III stars provided $\geq 50\%$ of the total amount of metals. 
    \\
    \item C-normal metal-poor halo stars have probably been imprinted mainly by Pop~II SNe, which provided $\geq 50\%$ of the total metals amount of their birth places. However, we might also find C-normal metal-poor stars which are pure descendants of the most energetic Pop~III SNe (hypernovae and PISN) with peculiar and outlier abundance patterns.
    
\end{itemize}

A key diagnostic employed to understand the origin of C-normal stars is the dispersion between the chemical abundances predicted by different models and its variation with respect to the pollution level from Pop~II stars. Indeed, the scatter between the abundances of different descendants decreases as the contribution of Pop~II stars to the metal pollution increases (see Sect.~\ref{sec:scatter}). If compared to the star-to-star dispersion of the chemical abundances of CEMP-no and C-normal halo stars, this supports the scenario where at $\rm[C/Fe]\sim 0$ the probability for metal-poor halo stars to be predominantly polluted by Pop~II stars is extremely high. Very recently, it has been shown that also the abundance dispersion of some high-redshift gaseous absorption systems increases with redshift (see Sodini et al. 2023, in prep.), denoting a possible trace left by Pop~III stars at $z \gtrsim 4$. 

Our new model provides a useful tool to analyse the abundances of metal-poor environments (present-day stars and high-redshift gas clouds) and to identify which of them has been likely enriched by the first Pop~III SNe and at which level.
This will soon become extremely important, when we will be able to exploit the chemical abundances of a huge amount of present-day stars provided by the 4MOST surveys \citep[see][]{Feltzing2018,Christlieb4MOST2019,deJong2019,Skuladottir2023dwarfs} for the stellar halo and the dwarf galaxies satellites of the Milky Way, and by WEAVE \citep[see][]{Dalton2014,Jin2023}, which will complement the work done by Gaia observing the entire Milky Way. Our model and new diagnostics have been already used to understand the origin of the recently observed high-z absorption systems \citep[see][Sodini et al. 2023, in prep]{Saccardi2023, Salvadori2023} and it will be of fundamental importance to guide future observations of high-z absorbers with ANDES on the ELT \citep[see][]{ANDES2022}, which aim at unveiling the signature of Pop~III SNe.

\section*{Acknowledgements}

The authors thank Marco Limongi for the inspiring discussions about the stellar yields. This project has received funding from the European Research Council (ERC) under the European Union’s Horizon 2020 research and innovation programme (grant agreement No 804240). I.V. and S.S. acknowledge support from the PRIN-MIUR17, prot. n. 2017T4ARJ5. 

%%%%%%%%%%%%%%%%%%%%%%%%%%%%%%%%%%%%%%%%%%%%%%%%%%
\section*{Data Availability}

The \citet{Heger2010} yields and the routines to use them are available at https://pypi.org/project/starfit/ and https://2sn.org/starfit/.



\bibliographystyle{mnras}
\bibliography{main} % if your bibtex file is called example.bib

%\begin{thebibliography}{99}
%Author A. N., 2013, Journal of Improbable Astronomy, 1, 1
%\end{thebibliography}

%%%%%%%%%%%%%%%%%%%%%%%%%%%%%%%%%%%%%%%%%%%%%%%%%%

%%%%%%%%%%%%%%%%% APPENDICES %%%%%%%%%%%%%%%%%%%%%

\appendix

\section{Direct comparison between Pop~III and Pop~II yields}
\label{sec:comparison_III_II}

\begin{figure*}
    \centering
    \includegraphics[width=\linewidth]{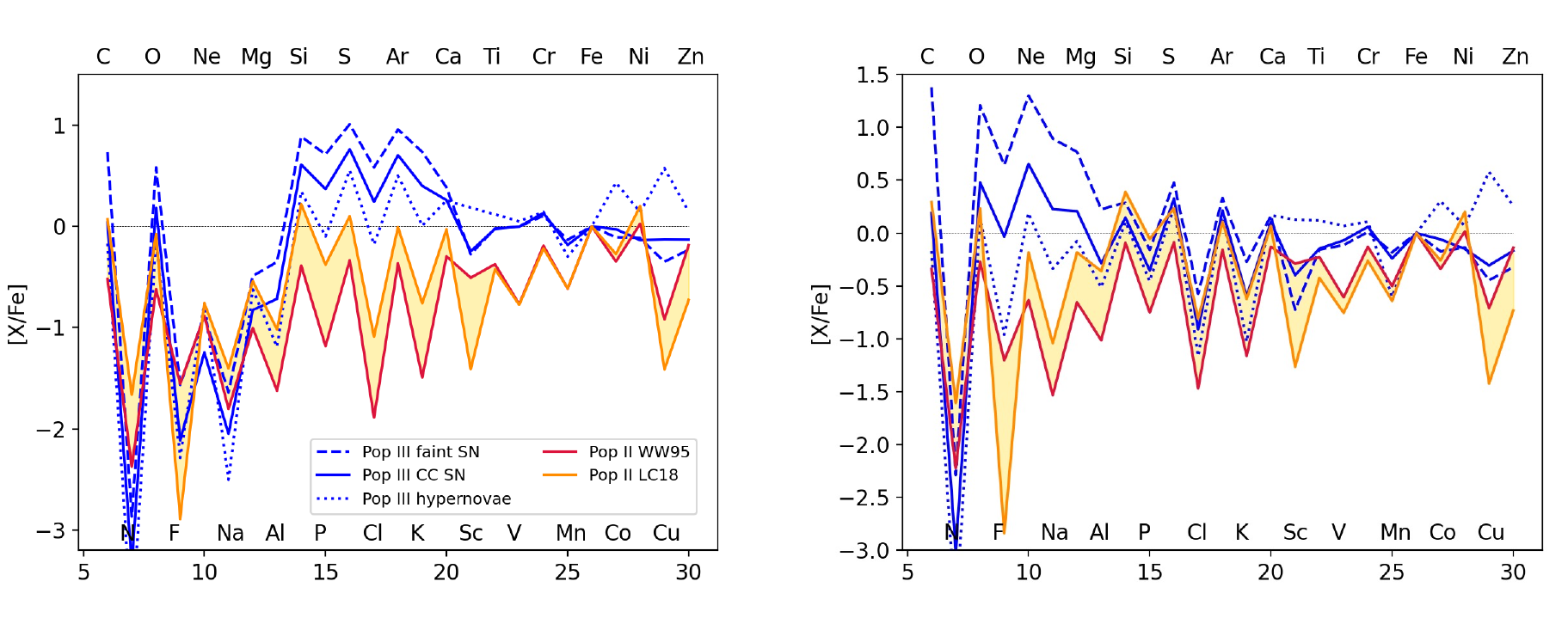}
    \caption{\reply{Abundance patterns, [X/Fe], for a star with 13\,M$_{\odot}$ (left) and 15\,M$_{\odot}$ (right), for the different yields adopted in the paper. \textit{Pop~III yields (blue):} faint SN with E$_{\rm SN}=0.6\times 10^{51}$ erg (dashed); core-collapse SN with E$_{\rm SN}=1.2\times 10^{51}$ erg (solid); hypernova with E$_{\rm SN}=10\times 10^{51}$ erg (dotted). All Pop~III yields are from \citealp[][]{Heger2010}, assuming mixing parameter $f_{\rm mix}=0.063$. \textit{Pop~II yields:}  core-collapse SN with E$_{\rm SN}\sim1.0\times 10^{51}$~erg from two sources; red is from \citealt[][]{Woosley1995}, and orange is from \citealp[][]{Limongi2018}. The yellow shaded area shows the intrinsic differences between the Pop~II yields from the two different groups.}}
    \label{fig:comparison_III_II}
\end{figure*}

\reply{The adopted chemical yields of Pop~III and Pop~II stars are intrinsically different. We show in Fig.~\ref{fig:comparison_III_II} a comparison between the yields of Pop~III stars (blue lines) and Pop~II stars (red from \citealt{Woosley1995} and orange from \citealt{Limongi2018}), adopted in this paper, for a 13 $\rm M_{\odot}$ (left) and 15 $\rm M_{\odot}$ star (right). First of all, Pop~III SNe can explode with a variety of energies: faint SNe (dashed) eject in the gas high quantities of the elements lighter than Ca, while hypernovae (dotted) are prolific of Zn. Unfortunately, in the adopted yields of Pop~II stars, there are no models exploding as faint SNe or hypernovae, therefore it is not possible to make a direct comparison. However, in Fig.~\ref{fig:comparison_III_II} we show that the yields of Pop~III (solid blue) and Pop~II (orange and red) core-collapse SNe are different. Depending on the initial mass of the star, the abundance patterns produced by Pop~III and Pop~II core-collapse SNe diverge for different chemical elements.} 

\reply{In Fig.~\ref{fig:comparison_III_II} the shaded yellow area represents the intrinsic uncertainty in the yields computation. Indeed, when the yields are computed by different groups, the discrepancies in [X/Fe] can be of the order of 1~dex.}

\section{The other chemical abundances of Pop~III descendants with different $f_{\rm PopIII}$}

\begin{figure*}
    \centering
    \includegraphics[width=0.9\textwidth]{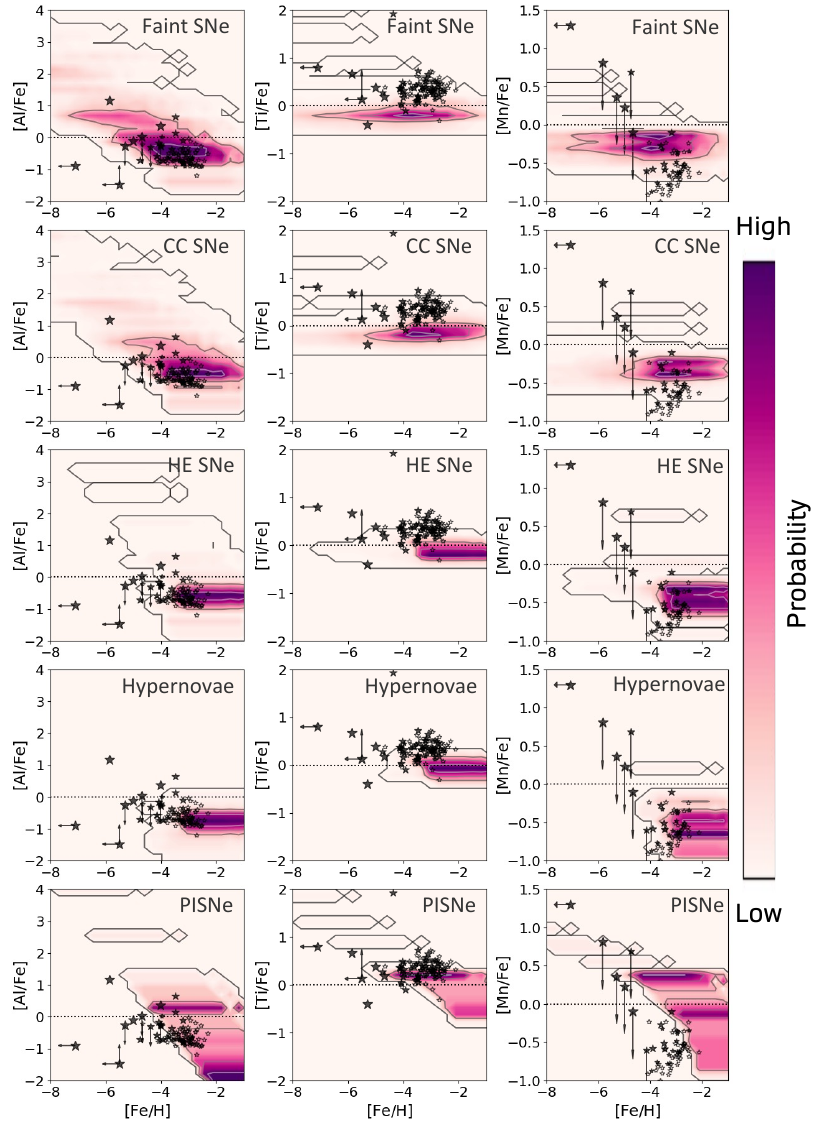}
    \caption{The same as Figs.~\ref{fig:PopIII_delta_1} and \ref{fig:PopIII_delta_2}, but for aluminium, titanium and manganese. Star symbols are observed chemical abundances of CEMP-no (filled, the sizes are proportional to the [C/Fe] values) and C-normal (open) halo star. Other elements are shown in Figs.~\ref{fig:PopIII_delta_1} (C, O, Mg) and \ref{fig:PopIII_delta_2} (Si, Ca, Zn).} 
    \label{fig:PopIII_delta_3}
\end{figure*}

\begin{figure*}
    \centering
    \includegraphics[width=\textwidth]{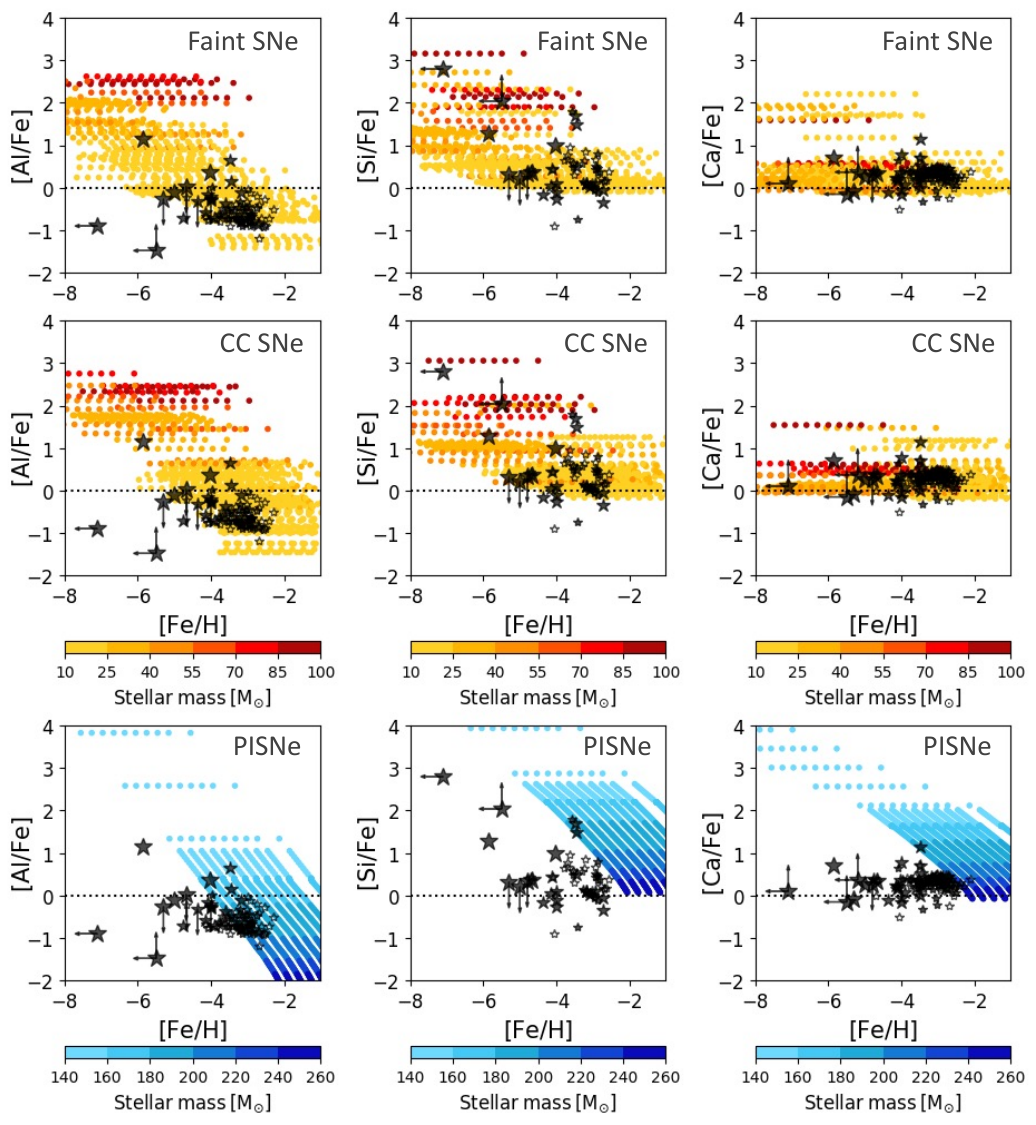}
    \caption{The same as Fig.~\ref{fig:PopIII_masses_1}, but for aluminium, silicon and calcium. Star symbols are observed chemical abundances of CEMP-no (filled, the sizes are proportional to the [C/Fe] values) and C-normal (open) halo star. Other elements are shown in Figs.~\ref{fig:PopIII_masses_1} (C, O, Mg) and \ref{fig:PopIII_masses_3} (Ti, Mn, Zn).} 
    \label{fig:PopIII_masses_2}
\end{figure*}

\begin{figure*}
    \centering
    \includegraphics[width=\textwidth]{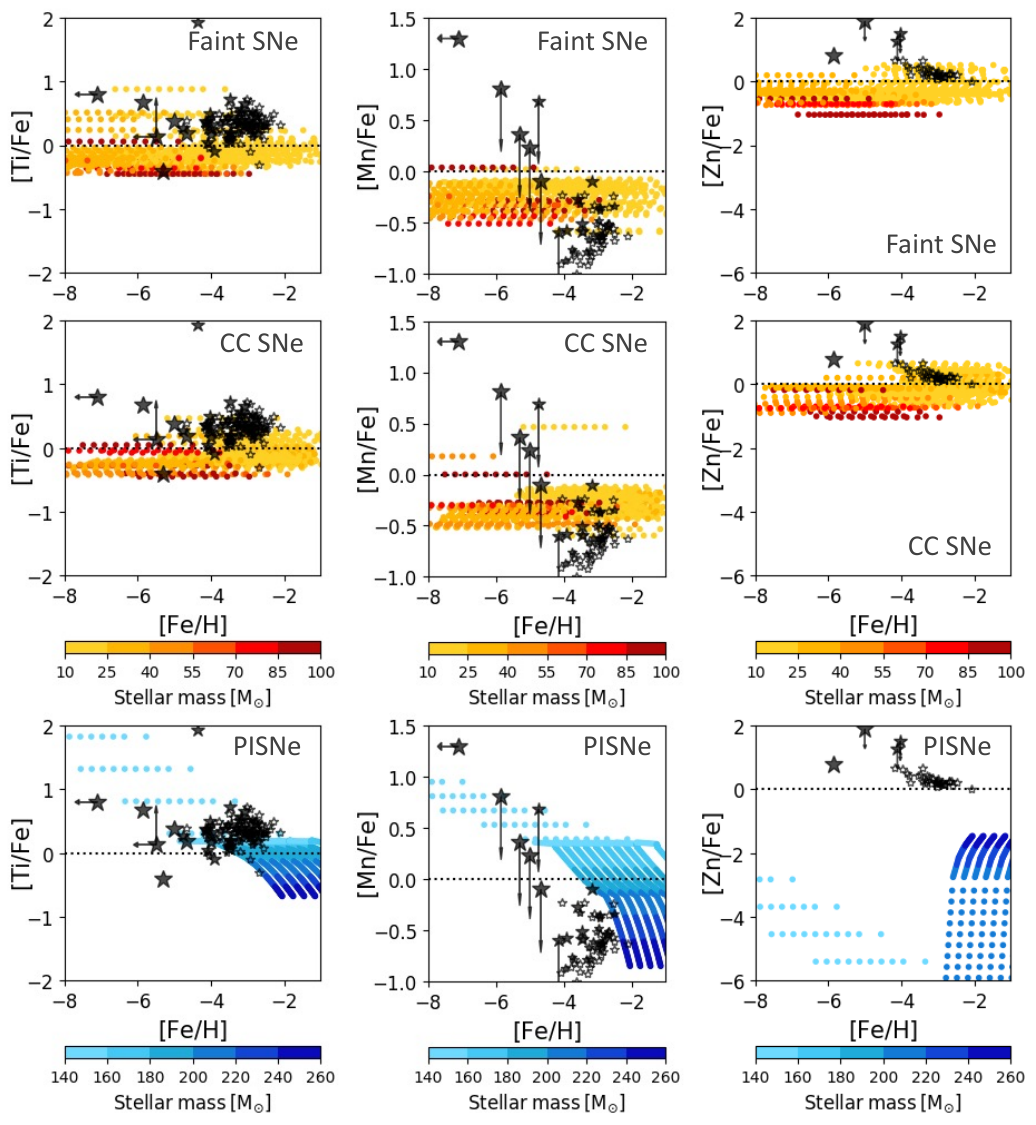}
    \caption{The same as Figs.~\ref{fig:PopIII_masses_1} and \ref{fig:PopIII_masses_2}, but for titanium, manganese, zinc. Star symbols are observed chemical abundances of CEMP-no (filled, the sizes are proportional to the [C/Fe] values) and C-normal (open) halo star. Other elements are shown in Figs.~\ref{fig:PopIII_masses_1} (C, O, Mg) and \ref{fig:PopIII_masses_2} (Al, Si, Ca).}
    \label{fig:PopIII_masses_3}
\end{figure*}

\begin{figure*}
    \centering
    \includegraphics[width=0.9\textwidth]{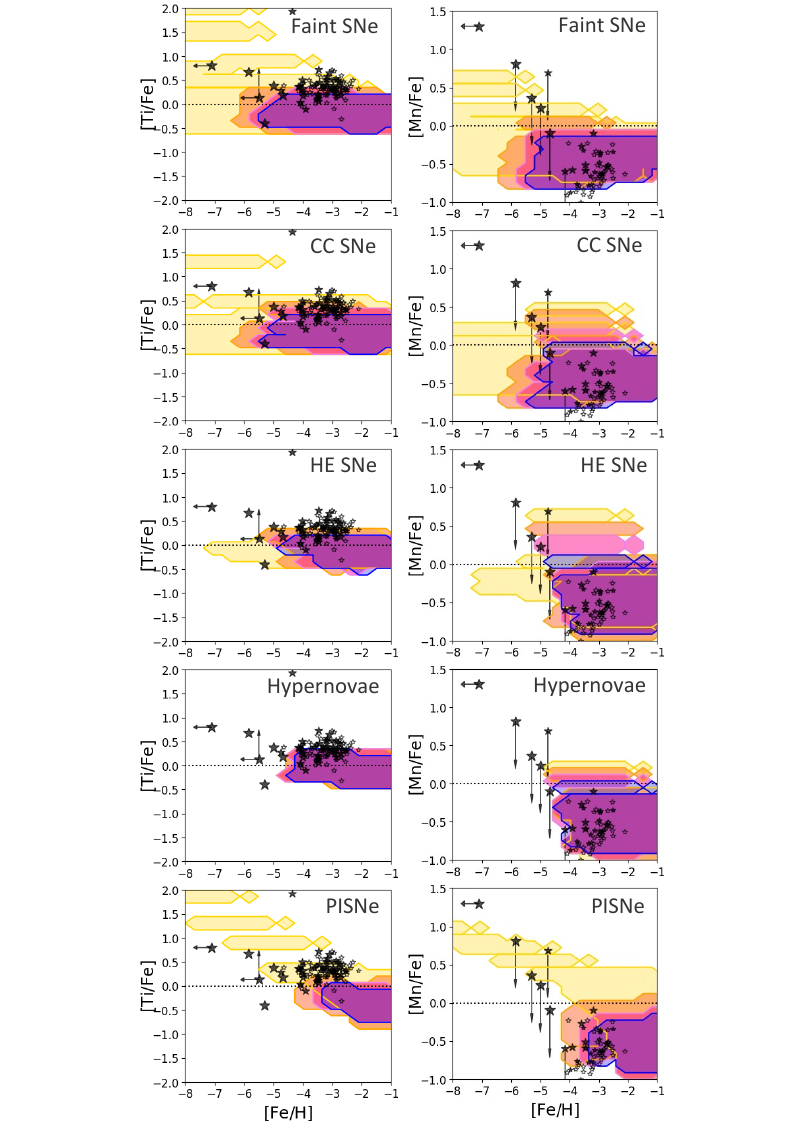}
    \caption{The same as Figs.~\ref{fig:PopII_1} and \ref{fig:PopII_2} but for titanium and manganese. Star symbols are observed chemical abundances of CEMP-no (filled, the sizes are proportional to the [C/Fe] values) and C-normal (open) halo star. Other elements are shown in Figs.~\ref{fig:PopII_1} (O, Mg, Si) and \ref{fig:PopII_2} (Al, Ca, Zn).}
    \label{fig:PopII_3}
\end{figure*}

In this Section we present the comparison between the abundance ratios predicted by our model for the descendants of Pop~III stars and the ones measured in the literature sample stars, which are not included in Sect.~\ref{results}. In the Figs.~\ref{fig:PopIII_delta_3}, \ref{fig:PopIII_masses_2} and \ref{fig:PopIII_masses_3} we show the abundances predicted for 100\% Pop~III descendants, in the Figs.~\ref{fig:PopII_2} and \ref{fig:PopII_3} the abundances predicted for the descendants contaminated also by normal Pop~II stars.

\label{sec:appendix_1}

\section{Exploring all the parameters}

\label{sec:appendix_2}

\begin{figure}
    \centering
    \includegraphics[width=\linewidth]{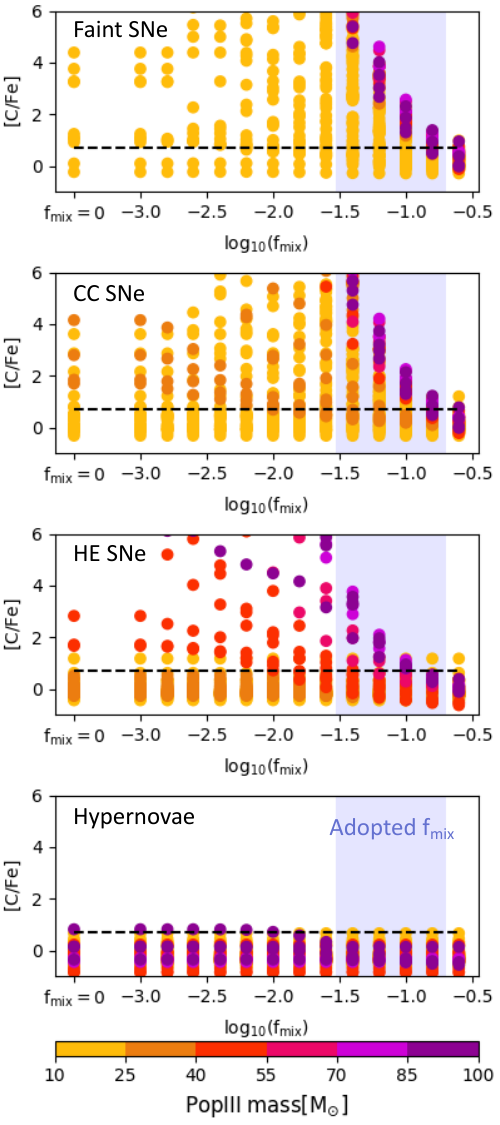}
    \caption{Abundance ratios of pure Pop~III descendants, [C/Fe], with respect to the logarithm of the mixing efficiency of a single Pop~III star for different initial stellar masses and SN explosion energies. In each panel one point represents the abundances for one set of parameters from \citet{Heger2010}. From the top to the bottom panel the Pop~III SN explosion energy increases: 0.6 (faint SNe), 1.2 (core-collapse SNe), 3.0 (high-energy SNe) and 10.0 (hypernovae) $\times 10^{51}$~erg. The points are colored depending on the initial mass of the progenitor, from yellow for $\sim 10-25 \mathrm{M_{\odot}}$, to purple for $\sim 85-100 \mathrm{M_{\odot}}$. The shaded area identifies the mixing efficiencies we adopted in this paper ($f_{\rm mix}=0.039, 0.063, 0.100, 0.158$).}
    \label{fig:mixing}
\end{figure}

We adopted the yields for primordial stars with masses~$\in [10, 100] \mathrm{M_{\odot}}$ from \citet{Heger2010}. The free parameters of their evolutionary model are the initial mass (120 stellar masses), the SN explosion energy (10 energies) of the star and the mixing efficiency of the stellar interior (14 fractions). After the SN explosion, a portion of the envelope falls back on the stellar remnant and doesn't get ejected in the ISM, thus not contributing to the pollution of the surrounding primordial gas. If the stellar interior is not mixed, a large part of the chemical elements produced in the most internal layers, i.e. the heaviest metals up to Zn, fall on the remnant after the SN explosion, leading to small amounts of Fe in the ISM and to high [C, O, Mg/Fe] abundance ratios in the ISM. The amount of the envelope that falls back on the remnant depends on the SN explosion energy and the mass of the star. Low-energy SNe succeed in ejecting smaller amounts of metals with respect to high-energy SNe, as well as low-mass stars with respect to high-mass stars.

Nonetheless, the stellar yields depend also on the mixing efficiency of the stellar interior. In Fig.~\ref{fig:mixing} we show the [C/Fe] ratio for single 100\% Pop~III descendants varying the mixing fraction, $f_{\rm mix}$, and the initial mass of the progenitor star. From the top to the bottom panel the SN explosion energy increases. In each panel we highlighted with a shaded area the adopted mixing efficiencies, $f_{\rm mix}=0.039, 0.063, 0.100, 0.158$, which roughly correspond to a boxcar of width equal to 4\% to 16\% of the helium core moving through the star. We chose to vary $f_{\rm mix}$ in this range because $f_{\rm mix}=0.100$ is the "standard mixing" suggested by \citet{Heger2010} and the [C/Fe] values of the faint SN descendants are in agreement with the ones observed for the most C-enhanced halo stars \citep[see][]{Iwamoto2005a,Keller}. For $f_{\rm mix}$ low enough ($\leq 0.039$), the descendants of Pop~III SNe which explode with $E_{\rm SN} \leq 3 \times 10^{51}$ erg, in the first three panels, can have [C/Fe] $> 6$. In these three cases the [C/Fe] ratio of the descendants tend to increase with the mass of the progenitor, as shown in Fig.~\ref{fig:PopIII_masses_1}. The [C/Fe] values of the descendants of Pop~III hypernovae, on the other hand, are not affected by the mixing efficiency. Indeed, the hypernovae are so energetic that they expel almost all the envelope and, even when the mixing efficiency is 0, the amount of iron ejected is almost the same as carbon. \reply{With $f_{\rm mix} < 0.039$, high-energy SNe have descendants with [C/Fe]$>+2.5$, yet with very low probabilities. In that case, we would still conclude that most C-enhanced halo stars are second generation stars, but, they could have also high-energy SNe progenitors.}

%%%%%%%%%%%%%%%%%%%%%%%%%%%%%%%%%%%%%%%%%%%%%%%%%%

% Don't change these lines
\bsp	% typesetting comment
\label{lastpage}
\end{document}